\documentclass[reprint,twocolumn,floatfix,altaffilletter,superscriptaddress,preprintnumbers,
tightenlines,showpacs,showkeys,nofootinbib,notoccite]{revtex4-2}

\usepackage[utf8]{inputenc}
\usepackage[colorlinks=true,citecolor=blue,linkcolor=blue]{hyperref}
\usepackage[normalem]{ulem}
\usepackage{amsmath,amssymb,mathrsfs,esvect}
\usepackage{epsfig}
\usepackage{graphicx}               
\usepackage{url}
\usepackage{color}
\usepackage{slashed}
\usepackage{multirow}
\usepackage{placeins}
\usepackage[dvipsnames]{xcolor}
\usepackage{epstopdf}
\usepackage{soul}
\usepackage{tikz}
\usepackage[capitalise, english]{cleveref}
\usepackage{siunitx}
\usepackage{xspace}
\usepackage{booktabs}
\usepackage[capitalise]{cleveref}
\usepackage{dcolumn}
\usepackage{bm}
\usepackage{orcidlink}
\usepackage{comment}
\usepackage{lineno}
\usepackage{marvosym}
\makeatletter
\renewcommand\bibsection{%
  \begingroup
  \let\MakeUppercase\relax  
  \let\uppercase\relax       
  \section*{References}      
  \endgroup
}
\makeatother

\hypersetup{
    colorlinks=true,
    linkcolor=black,
    citecolor=blue,
}

\renewcommand{\figurename}{{\bf Fig.}}

\usetikzlibrary{trees}
\usetikzlibrary{decorations.pathmorphing}
\usetikzlibrary{decorations.markings}

\allowdisplaybreaks
\begin{document}

\title{Dynamical Dark Energy in light of the DESI DR2 Baryonic Acoustic Oscillations Measurements}

\author{Gan Gu\orcidlink{0009-0007-9215-489X}}
\thanks{These authors contributed equally to this work.}
\affiliation{National Astronomical Observatories, Chinese Academy of Sciences, Beijing, 100101, P.R.China}
\affiliation{University of Chinese Academy of Sciences, Beijing, 100049, P.R.China}

\author{Xiaoma Wang\orcidlink{0000-0003-0216-1230}}
\thanks{These authors contributed equally to this work.}
\affiliation{National Astronomical Observatories, Chinese Academy of Sciences, Beijing, 100101, P.R.China}
\affiliation{University of Chinese Academy of Sciences, Beijing, 100049, P.R.China}

\author{Yuting Wang\orcidlink{0000-0001-7756-8479}\textsuperscript{\Letter}}

\affiliation{National Astronomical Observatories, Chinese Academy of Sciences, Beijing, 100101, P.R.China}
\affiliation{Institute for Frontiers in Astronomy and Astrophysics, Beijing Normal University, Beijing, 102206, P.R.China}

\author{Gong-Bo Zhao\orcidlink{0000-0003-4726-6714}\textsuperscript{\Letter}}

\affiliation{National Astronomical Observatories, Chinese Academy of Sciences, Beijing, 100101, P.R.China}
\affiliation{University of Chinese Academy of Sciences, Beijing, 100049, P.R.China}
\affiliation{Institute for Frontiers in Astronomy and Astrophysics, Beijing Normal University, Beijing, 102206, P.R.China}

\author{Levon Pogosian\orcidlink{0000-0001-5108-0854}\textsuperscript{\Letter}}

\affiliation{Department of Physics, Simon Fraser University, Burnaby, BC, V5A 1S6, Canada}

\author{Kazuya~Koyama\orcidlink{0000-0001-6727-6915}\textsuperscript{\Letter}}

\affiliation{Institute of Cosmology \& Gravitation, University of Portsmouth, Portsmouth, PO1 3FX, UK}
\affiliation{
Kavli IPMU (WPI), UTIAS, The University of Tokyo, Kashiwa, Chiba 277-8583, Japan}

\author{John~A.~Peacock\orcidlink{0000-0002-1168-8299}}
\affiliation{Institute for Astronomy, University of Edinburgh, Royal Observatory, Blackford Hill, Edinburgh EH9 3HJ, UK}

\author{Zheng Cai\orcidlink{0000-0001-8467-6478}}
\affiliation{Department of Astronomy, Tsinghua University, Beijing 100084, P.R.China}

\author{Jorge L. Cervantes-Cota\orcidlink{0000-0002-3057-6786}}
\affiliation{Departamento de Física, Instituto Nacional de Investigaciones Nucleares, Apartado Postal 18-1027, Col. Escandón, Ciudad de México, 11801, México}

\author{Mustapha Ishak\orcidlink{0000-0002-6024-466X}}
\affiliation{Department of Physics, The University of Texas at Dallas, 800 W. Campbell Rd., Richardson, TX 75080, USA}

\author{Arman Shafieloo\orcidlink{0000-0001-6815-0337}}
\affiliation{Korea Astronomy and Space Science Institute, 776, Daedeokdae-ro, Yuseong-gu, Daejeon 34055, Republic of Korea}
\affiliation{University of Science and Technology, 217 Gajeong-ro, Yuseong-gu, Daejeon 34113, Republic of Korea}

\author{Ruiyang Zhao\orcidlink{0000-0002-7284-7265}}
\affiliation{National Astronomical Observatories, Chinese Academy of Sciences, Beijing, 100101, P.R.China}
\affiliation{University of Chinese Academy of Sciences, Beijing, 100049, P.R.China}

\author{Steven Ahlen\orcidlink{0000-0001-6098-7247}}
\affiliation{Physics Department, Boston University, 590 Commonwealth Avenue, Boston, MA 02215, USA}

\author{Davide Bianchi\orcidlink{0000-0001-9712-0006}}
\affiliation{Dipartimento di Fisica ``Aldo Pontremoli'', Universit\`a degli Studi di Milano, Via Celoria 16, I-20133 Milano, Italy}
\affiliation{INAF-Osservatorio Astronomico di Brera, Via Brera 28, 20122 Milano, Italy}

\author{David Brooks\orcidlink{0000-0002-8458-5047}}
\affiliation{Department of Physics \& Astronomy, University College London, Gower Street, London, WC1E 6BT, UK}

\author{Todd Claybaugh\orcidlink{0000-0002-5024-6987}}
\affiliation{Lawrence Berkeley National Laboratory, 1 Cyclotron Road, Berkeley, CA 94720, USA}

\author{Shaun Cole\orcidlink{0000-0002-5954-7903}}
\affiliation{Institute for Computational Cosmology, Department of Physics, Durham University, South Road, Durham DH1 3LE, UK}

\author{Axel de la Macorra\orcidlink{0000-0002-1769-1640}}
\affiliation{Instituto de F\'{\i}sica, Universidad Nacional Aut\'{o}noma de M\'{e}xico,  Circuito de la Investigaci\'{o}n Cient\'{\i}fica, Ciudad Universitaria, Cd. de M\'{e}xico  C.~P.~04510,  M\'{e}xico}

\author{Arnaud de Mattia\orcidlink{0000-0003-0920-2947}}
\affiliation{IRFU, CEA, Universit\'{e} Paris-Saclay, F-91191 Gif-sur-Yvette, France}

\author{Peter Doel\orcidlink{0000-0002-6397-4457}}
\affiliation{Department of Physics \& Astronomy, University College London, Gower Street, London, WC1E 6BT, UK}

\author{Simone~Ferraro\orcidlink{0000-0003-4992-7854}}
\affiliation{Lawrence Berkeley National Laboratory, 1 Cyclotron Road, Berkeley, CA 94720, USA}
\affiliation{University of California, Berkeley, 110 Sproul Hall \#5800 Berkeley, CA 94720, USA}

\author{Jaime E. Forero-Romero\orcidlink{0000-0002-2890-3725}}
\affiliation{Departamento de F\'isica, Universidad de los Andes, Cra. 1 No. 18A-10, Edificio Ip, CP 111711, Bogot\'a, Colombia}
\affiliation{Observatorio Astron\'omico, Universidad de los Andes, Cra. 1 No. 18A-10, Edificio H, CP 111711 Bogot\'a, Colombia}

\author{Enrique Gaztañaga\orcidlink{0000-0001-9632-0815}}
\affiliation{Institut d'Estudis Espacials de Catalunya (IEEC), c/ Esteve Terradas 1, Edifici RDIT, Campus PMT-UPC, 08860 Castelldefels, Spain}
\affiliation{Institute of Cosmology \& Gravitation, University of Portsmouth, Portsmouth, PO1 3FX, UK}
\affiliation{Institute of Space Sciences, ICE-CSIC, Campus UAB, Carrer de Can Magrans s/n, 08913 Bellaterra, Barcelona, Spain}

\author{Satya Gontcho A Gontcho\orcidlink{0000-0003-3142-233X}}
\affiliation{Lawrence Berkeley National Laboratory, 1 Cyclotron Road, Berkeley, CA 94720, USA}

\author{Gaston Gutierrez\orcidlink{0000-0003-0825-0517}}
\affiliation{Fermi National Accelerator Laboratory, PO Box 500, Batavia, IL 60510, USA}

\author{ChangHoon Hahn\orcidlink{0000-0003-1197-0902}}
\affiliation{Steward Observatory, University of Arizona, 933 N. Cherry Avenue, Tucson, AZ 85721, USA}

\author{Cullan Howlett\orcidlink{0000-0002-1081-9410}}
\affiliation{School of Mathematics and Physics, University of Queensland, Brisbane, QLD 4072, Australia}

\author{Robert Kehoe}
\affiliation{Department of Physics, Southern Methodist University, 3215 Daniel Avenue, Dallas, TX 75275, USA}

\author{David Kirkby\orcidlink{0000-0002-8828-5463}}
\affiliation{Department of Physics and Astronomy, University of California, Irvine, 92697, USA}

\author{Jean-Paul Kneib\orcidlink{0000-0002-4616-4989}}
\affiliation{Institute of Physics, Laboratory of Astrophysics, \'Ecole Polytechnique F\'ed\'erale de Lausanne (EPFL), Observatoire de Sauverny, CH-1290 Versoix, Switzerland}
\affiliation{Aix Marseille Universit\'e, CNRS, LAM (Laboratoire d'Astrophysique de Marseille) UMR 7326, F13388, Marseille, France}

\author{Anthony Kremin\orcidlink{0000-0001-6356-7424}}
\affiliation{Lawrence Berkeley National Laboratory, 1 Cyclotron Road, Berkeley, CA 94720, USA}

\author{Ofer Lahav\orcidlink{0000-0002-1134-9035}}
\affiliation{Department of Physics \& Astronomy, University College London, Gower Street, London, WC1E 6BT, UK}

\author{Martin Landriau\orcidlink{0000-0003-1838-8528}}
\affiliation{Lawrence Berkeley National Laboratory, 1 Cyclotron Road, Berkeley, CA 94720, USA}

\author{Laurent Le Guillou\orcidlink{0000-0001-7178-8868}}
\affiliation{Sorbonne Universit\'{e}, CNRS/IN2P3, Laboratoire de Physique Nucl\'{e}aire et de Hautes Energies (LPNHE), FR-75005 Paris, France}

\author{Alexie~Leauthaud\orcidlink{0000-0002-3677-3617}}
\affiliation{Department of Astronomy and Astrophysics, UCO/Lick Observatory, University of California, 1156 High Street, Santa Cruz, CA 95064, USA}
\affiliation{Department of Astronomy and Astrophysics, University of California, Santa Cruz, 1156 High Street, Santa Cruz, CA 95065, USA}

\author{Michael Levi\orcidlink{0000-0003-1887-1018}}
\affiliation{Lawrence Berkeley National Laboratory, 1 Cyclotron Road, Berkeley, CA 94720, USA}

\author{Marc Manera\orcidlink{0000-0003-4962-8934}}
\affiliation{Departament de F\'{i}sica, Serra H\'{u}nter, Universitat Aut\`{o}noma de Barcelona, 08193 BELlaterra (Barcelona), Spain}
\affiliation{Institut de F\'{i}sica d’Altes Energies (IFAE), The Barcelona Institute of Science and Technology, Edifici Cn, Campus UAB, 08193, Bellaterra (Barcelona), Spain}

\author{Aaron Meisner\orcidlink{0000-0002-1125-7384}}
\affiliation{NSF NOIRLab, 950 N. Cherry Ave., Tucson, AZ 85719, USA}

\author{Ramon Miquel\orcidlink{0000-0002-6610-4836}}
\affiliation{Instituci\'{o} Catalana de Recerca i Estudis Avan\c{c}ats, Passeig de Llu\'{\i}s Companys, 23, 08010 Barcelona, Spain}
\affiliation{Institut de F\'{i}sica d’Altes Energies (IFAE), The Barcelona Institute of Science and Technology, Edifici Cn, Campus UAB, 08193, Bellaterra (Barcelona), Spain}

\author{John Moustakas\orcidlink{0000-0002-2733-4559}}
\affiliation{Department of Physics and Astronomy, Siena College, 515 Loudon Road, Loudonville, NY 12211, USA}

\author{Andrea Muñoz-Gutiérrez}
\affiliation{Instituto de F\'{\i}sica, Universidad Nacional Aut\'{o}noma de M\'{e}xico,  Circuito de la Investigaci\'{o}n Cient\'{\i}fica, Ciudad Universitaria, Cd. de M\'{e}xico  C.~P.~04510,  M\'{e}xico}

\author{Seshadri Nadathur\orcidlink{0000-0001-9070-3102}}
\affiliation{Institute of Cosmology \& Gravitation, University of Portsmouth, Portsmouth, PO1 3FX, UK}

\author{Jeffrey A. Newman\orcidlink{0000-0001-8684-2222}}
\affiliation{Department of Physics \& Astronomy and Pittsburgh Particle Physics, Astrophysics, and Cosmology Center (PITT PACC), University of Pittsburgh, 3941 O'Hara Street, Pittsburgh, PA 15260, USA}

\author{Nathalie Palanque-Delabrouille\orcidlink{0000-0003-3188-784X}}
\affiliation{IRFU, CEA, Universit\'{e} Paris-Saclay, F-91191 Gif-sur-Yvette, France}
\affiliation{Lawrence Berkeley National Laboratory, 1 Cyclotron Road, Berkeley, CA 94720, USA}

\author{Will Percival\orcidlink{0000-0002-0644-5727}}
\affiliation{Department of Physics and Astronomy, University of Waterloo, 200 University Ave W, Waterloo, ON N2L 3G1, Canada}
\affiliation{Perimeter Institute for Theoretical Physics, 31 Caroline St. North, Waterloo, ON N2L 2Y5, Canada}
\affiliation{Waterloo Centre for Astrophysics, University of Waterloo, 200 University Ave W, Waterloo, ON N2L 3G1, Canada}

\author{Francisco Prada\orcidlink{0000-0001-7145-8674}}
\affiliation{Instituto de Astrof\'{i}sica de Andaluc\'{i}a (CSIC), Glorieta de la Astronom\'{i}a, s/n, E-18008 Granada, Spain}

\author{Ignasi P\'erez-R\`afols\orcidlink{0000-0001-6979-0125}}
\affiliation{Departament de F\'isica, EEBE, Universitat Polit\`ecnica de Catalunya, c/Eduard Maristany 10, 08930 Barcelona, Spain}

\author{Graziano Rossi}
\affiliation{Department of Physics and Astronomy, Sejong University, 209 Neungdong-ro, Gwangjin-gu, Seoul 05006, Republic of Korea}

\author{Lado Samushia\orcidlink{0000-0002-1609-5687}}
\affiliation{Abastumani Astrophysical Observatory, Tbilisi, GE-0179, Georgia}
\affiliation{Department of Physics, Kansas State University, 116 Cardwell Hall, Manhattan, KS 66506, USA}
\affiliation{Faculty of Natural Sciences and Medicine, Ilia State University, 0194 Tbilisi, Georgia}

\author{Eusebio Sanchez\orcidlink{0000-0002-9646-8198}}
\affiliation{CIEMAT, Avenida Complutense 40, E-28040 Madrid, Spain}

\author{David Schlegel\orcidlink{0000-0002-5042-5088}}
\affiliation{Lawrence Berkeley National Laboratory, 1 Cyclotron Road, Berkeley, CA 94720, USA}

\author{Hee-Jong Seo\orcidlink{0000-0002-6588-3508}}
\affiliation{Department of Physics \& Astronomy, Ohio University, 139 University Terrace, Athens, OH 45701, USA}

\author{David Sprayberry}
\affiliation{NSF NOIRLab, 950 N. Cherry Ave., Tucson, AZ 85719, USA}

\author{Gregory Tarl\'{e}\orcidlink{0000-0003-1704-0781}}
\affiliation{Department of Physics, University of Michigan, 450 Church Street, Ann Arbor, MI 48109, USA}

\author{Michael Walther\orcidlink{0000-0002-1748-3745}}
\affiliation{University Observatory, Faculty of Physics, Ludwig-Maximilians-Universit\"{a}t, Scheinerstr. 1, 81677 M\"{u}nchen, Germany}
\affiliation{Excellence Cluster ORIGINS, Boltzmannstrasse 2, D-85748 Garching, Germany}

\author{Benjamin Alan Weaver}
\affiliation{NSF NOIRLab, 950 N. Cherry Ave., Tucson, AZ 85719, USA}

\author{Pauline Zarrouk\orcidlink{0000-0002-7305-9578}}
\affiliation{Sorbonne Universit\'{e}, CNRS/IN2P3, Laboratoire de Physique Nucl\'{e}aire et de Hautes Energies (LPNHE), FR-75005 Paris, France}

\author{Cheng~Zhao\orcidlink{0000-0002-1991-7295}}
\affiliation{Department of Astronomy, Tsinghua University, Beijing 100084, P.R.China}

\author{Rongpu~Zhou\orcidlink{0000-0001-5381-4372}}
\affiliation{Lawrence Berkeley National Laboratory, 1 Cyclotron Road, Berkeley, CA 94720, USA}

\author{Hu Zou\orcidlink{0000-0002-6684-3997}}
\affiliation{National Astronomical Observatories, Chinese Academy of Sciences, Beijing, 100101, P.R.China}

\date{\today}

\begin{abstract}
Understanding whether cosmic acceleration arises from a cosmological constant or a dynamical component is a central goal of cosmology, and the Dark Energy Spectroscopic Instrument (DESI) enables stringent tests with high-precision distance measurements. We analyze baryon acoustic oscillation (BAO) measurements from DESI Data Release 1 (DR1) and Data Release 2 (DR2), combined with Type Ia supernovae and a cosmic microwave background (CMB) distance prior. With the larger statistical power and wider redshift coverage of DR2, the preference for dynamical dark energy does not diminish relative to DR1. Using both a shape-function reconstruction and non-parametric approaches with a Horndeski-motivated correlation prior, we find that the dark-energy equation of state $w(z)$ varies with redshift. BAO data alone yield modest constraints, but in combination with independent supernova compilations and the CMB prior they strengthen the evidence for dynamics. Bayesian model comparison shows moderate support for departures from $\Lambda$CDM when multiple degrees of freedom in $w(z)$ are allowed, corresponding to $\approx3\sigma$ tension with $\Lambda$CDM (and higher for some data sets). Despite methodological differences, our results are consistent with companion DESI papers, underscoring the complementarity of approaches. Possible systematics remain under study; forthcoming DESI, \emph{Euclid}, and next-generation CMB data will provide decisive tests.
\end{abstract}

\maketitle

\begingroup
\renewcommand\thefootnote{\Letter}
\footnotetext{Correspondence: \\
\href{mailto:gbzhao@nao.cas.cn}{gbzhao@nao.cas.cn} \\
\href{mailto:levon@sfu.ca}{levon@sfu.ca} \\
\href{mailto:kazuya.koyama@port.ac.uk}{kazuya.koyama@port.ac.uk} \\
\href{mailto:ytwang@nao.cas.cn}{ytwang@nao.cas.cn}}
\endgroup

\section{Introduction}

One of the most enduring puzzles in modern cosmology is the accelerating expansion of the Universe. The impetus for introducing an additional component beyond ordinary matter first arose in the 1980s, when early cosmic microwave background (CMB) anisotropy calculations showed that a low-density open model would conflict with small-scale CMB observations. Subsequently, large-scale clustering analyses \citep{Efstathiou1990} and baryon-fraction arguments \citep{White1993} made it clear that a Universe consisting solely of critical-density dark matter was untenable, prompting the emergence of a cosmological constant ($\Lambda$) as the leading explanation. Observations of Type Ia supernovae (SNe) then provided direct evidence for an accelerating cosmic expansion \citep{SN98a,SN98b}, and by the turn of the millennium, the $\Lambda$CDM framework was firmly established as the consensus model of cosmology. This paradigm has been continuously supported by increasingly precise CMB measurements \citep{WMAP} and baryon acoustic oscillation (BAO) data \citep{eBOSS17}, solidifying its status as the standard picture of cosmic acceleration.

Despite this consensus, the fundamental cause of cosmic acceleration remains elusive. A widely favored interpretation involves postulating an effective dark energy (DE) component that dominates the current energy density of the Universe (see \cite{DErev1,DErev2,DErev3} for reviews). Provided its sound speed is not too small, DE can be treated as nearly homogeneous on sub-sound-horizon scales, leaving its main cosmological signatures determined by its energy density $\rho$ and pressure $P$. The equation of state (EoS), $w \equiv P/c^2 \rho$, serves as a crucial diagnostic: $w = -1$ corresponds to vacuum energy, while dynamical DE models allow $w$ to vary with time, as exemplified by quintessence \citep{quintessence}, phantom \citep{phantom}, quintom \citep{quintom}, $k$-essence \citep{k-essence}, Chaplygin gas \citep{Cgas}, and holographic dark energy \citep{HDE}. Subclasses of these scenarios can be further differentiated by examining their trajectories in the phase space spanned by $\bigl(w,\,\mathrm{d}w/\mathrm{d}\ln a\bigr)$, where $a$ is the cosmic scale factor \citep{CL05}.

Recent observational progress, including supernova measurements from the Dark Energy Survey (DES) \citep{DESSNY5}, as well as other large SNe compilations such as
PantheonPlus \citep{Scolnic:2021amr,Brout:2022} and
Union \cite{Rubin:2023}, high-precision CMB data from the Planck Collaboration \citep{Planck18}, and BAO results from the Sloan Digital Sky Survey (SDSS) eBOSS program \citep{eBOSSDR16}, has significantly refined our insight into DE. These complementary probes constrain DE-related parameters with growing precision, thus enhancing our understanding of cosmic acceleration. While combined analyses that integrate all available datasets provide the most stringent overall limits, it remains instructive to consider each probe in isolation. Such a modular approach not only underscores the unique strengths and systematics of each measurement but also furnishes critical cross-checks, reinforcing the robustness of DE constraints.

The Dark Energy Spectroscopic Instrument (DESI)'s Data Release~1 (DR1) \citep{DESI2024.I.DR1} and Data Release~2 (DR2) \citep{DESI.DR2.BAO.lya,DESI.DR2.BAO.cosmo} BAO observations place stringent limits on the Universe’s background expansion history, achieving a level of precision on a par with recent SNe surveys. These include the PantheonPlus compilation of 1{,}550 spectroscopically confirmed SNe \cite{Scolnic:2021amr,Brout:2022}, the Union3 compilation of 2{,}087 SNe \cite{Rubin:2023}, and the five-year Dark Energy Survey sample (DESY${5}$) \cite{DESSNY5}. 

In this paper we probe dark-energy dynamics along two complementary routes. First, we fit the distance ladder with a shape-function formalism that makes no explicit assumption about the equation of state. Secondly, we carry out a fully Bayesian reconstruction of $w(z)$ in which a correlation prior derived from the Horndeski gravity \cite{Horndeski:1974wa} yields an interpretable Bayes factor, recovering the standard $(w_0,w_a)$ result \cite{Chevallier2001,Linder2003} with two effective degrees of freedom. Although these techniques differ from the Gaussian-process and $(w_0,w_a)$ analyses presented in the companion DESI studies \cite{DESI.DR2.BAO.cosmo,Y3.cpe-s1.Lodha.2025}, all approaches paint a coherent picture: DESI DR1 and DR2 BAO measurements, together with three different Type Ia supernova samples, indicate the same mild, oscillatory departure from a cosmological constant. The convergence of results across data sets and methodologies strengthens the statistical case for an evolving dark energy component.

\begin{figure*}[htp]
    \centering
    \includegraphics[width=0.95\textwidth]{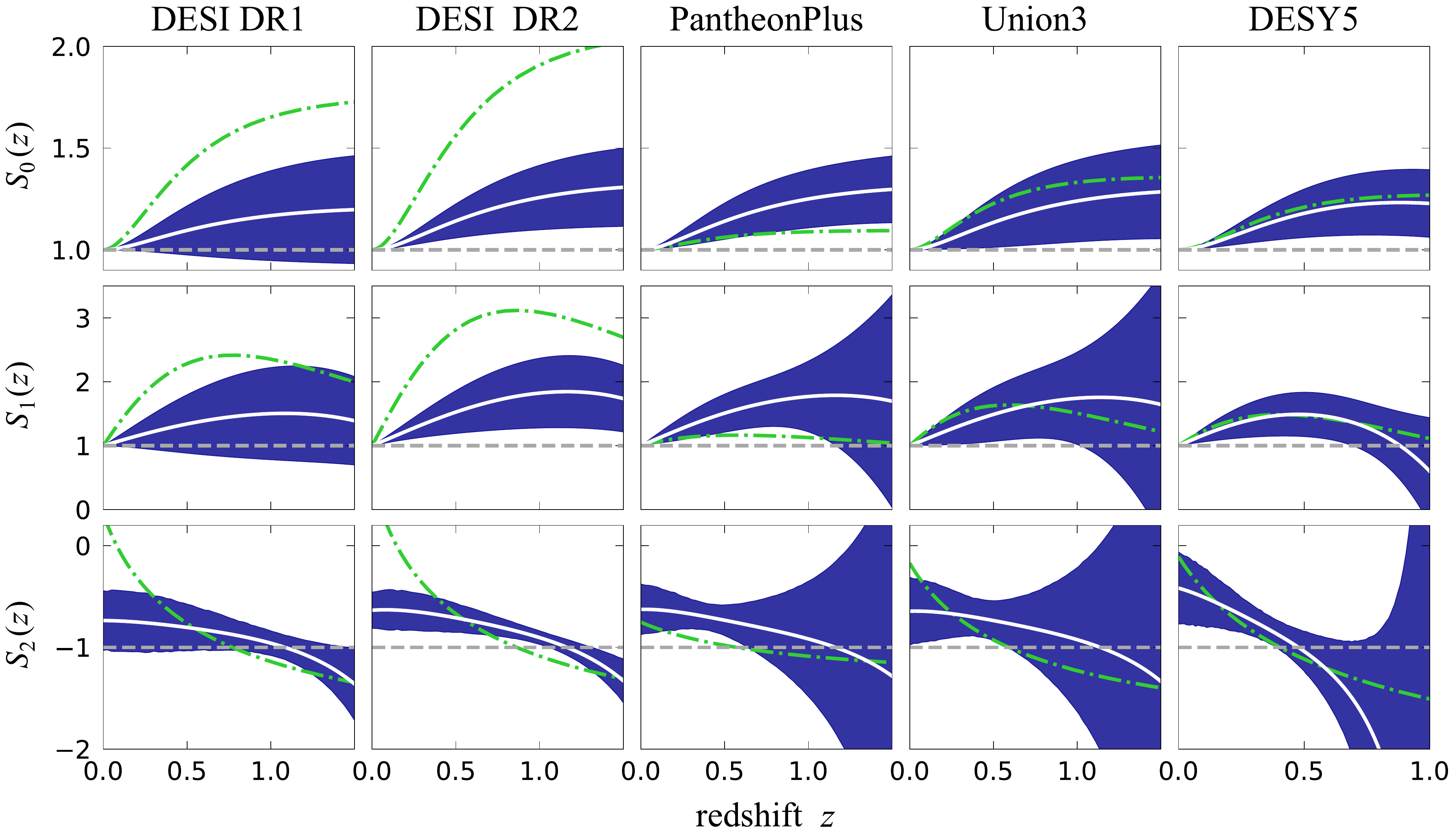}
    \caption{{\bf Reconstructed shape functions \(S_i\) \((i = 0, 1, 2)\) derived from multiple cosmic-distance measurements, displayed as a function of redshift \(z\).} All five distance probes show a consistent, $\Lambda$CDM-deviating trend in the $S$-functions, strongest in DESI DR2. Each column corresponds to a different dataset: DESI BAO DR~1 (leftmost), DESI BAO DR~2, PantheonPlus, Union3, and DESY5, (rightmost). From top to bottom, the panels show \(S_0(z)\), \(S_1(z)\), and \(S_2(z)\), as defined in Eqs. (\ref{eq:DEfunctions}), which capture the time evolution of the energy density of dark energy, the pressure of dark energy, and the equation of state of dark energy, respectively. In each panel, the solid white line indicates the best-fit reconstruction, while the blue shaded region denotes the 68\% confidence-level uncertainty, derived according to Eqs.~\eqref{eq:chi_H}--\eqref{eq:DEfunctions}. The green dash-dotted curve represents the best-fit shape function from the CPL parameterization, and the horizontal dashed line is the corresponding prediction of the \(\Lambda\)CDM model. See texts for further details.
}
    
    \label{fig:shape_functions}
\end{figure*}

\begin{figure*}[htp]
    \centering
    \includegraphics[width=0.9\textwidth]{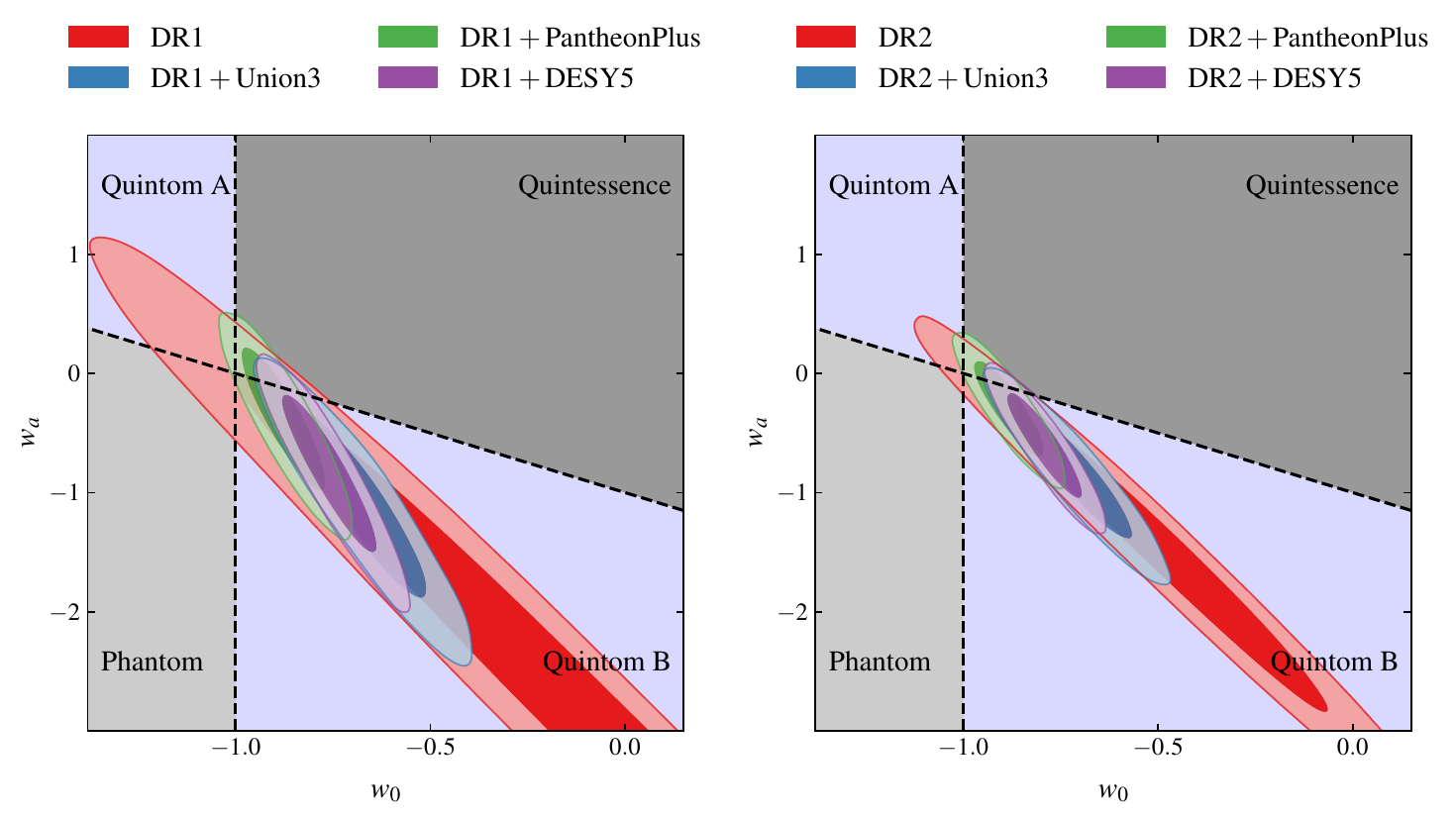}
    \caption{
        {\bf 68\% and 95\% credible intervals constraints on the dark energy equation-of-state parameters $(w_0, w_a)$ for various data combinations.}
        The left panel displays results based on DESI DR1 measurements, while the right panel shows DR2. 
        Different colors indicate distinct dataset combinations, as labeled. In all cases we assume in addition a BBN constraint on $\Omega_{\rm b}h^2$ and a CMB constraint on $\theta_*$. The parameter space is subdivided according to four dark energy models: 
        Quintom A ($w > -1$ in the past and $w < -1$ today), 
        Quintessence ($w > -1$ at all epochs), 
        Phantom ($w < -1$ at all epochs), 
        and Quintom B ($w < -1$ in the past but $w > -1$ today). 
        The intersection point of the two dashed lines corresponds to the 
    $\Lambda \mathrm{CDM}$ limit $(w_0,\,w_a)=(-1,0)$.}  
    
    \label{fig:w0wa}
\end{figure*}

\section{A shape-function analysis of dark energy}

\begin{figure*}[htp]
    \centering
    \includegraphics[width=\textwidth]{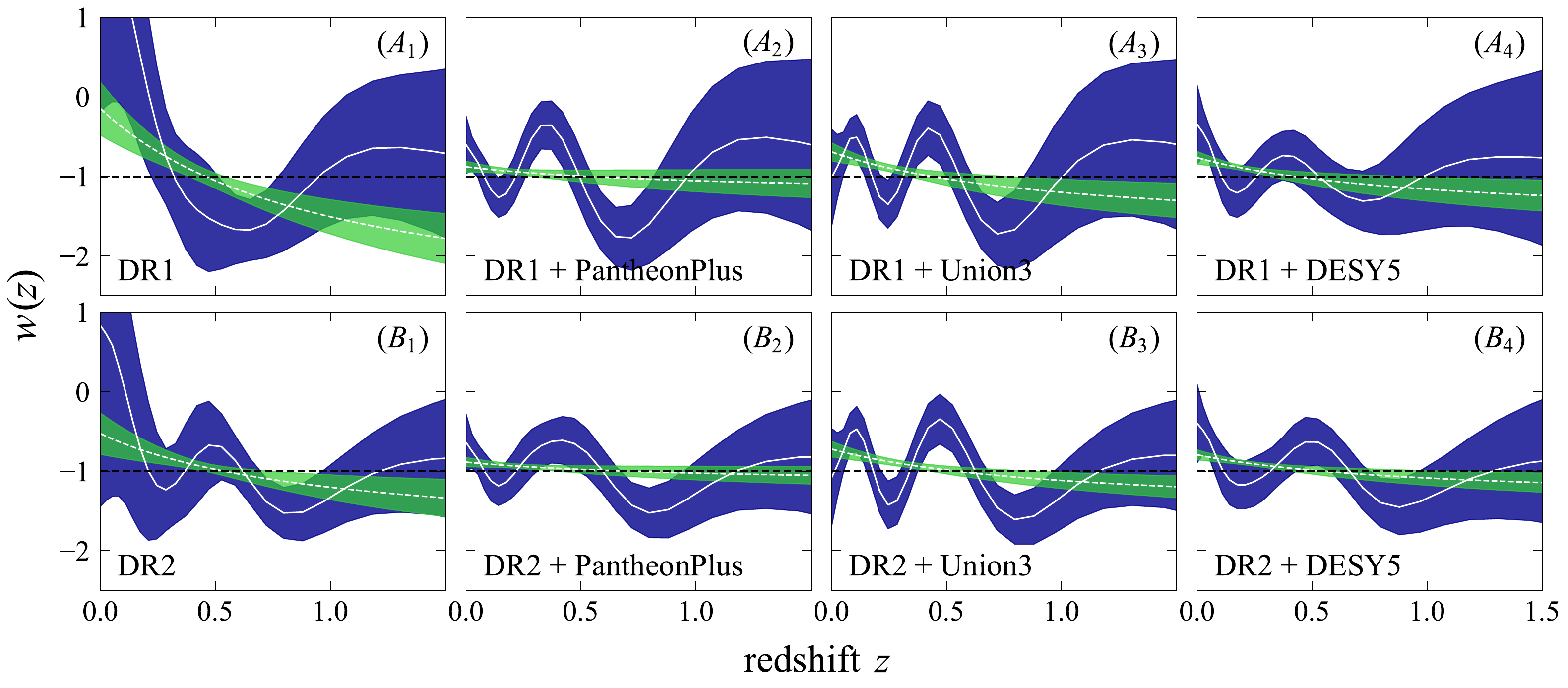}
    \caption{
        {\bf Reconstructed dark energy equation of state $w(z)$ from multiple datasets.}
        Panels $(A_1$--$A_4)$ show DESI DR1 BAO results, both alone and in combination with three independent Type~Ia supernova (SNe) datasets, under two different approaches: 
        the correlation-prior method (bottom-layered, dark-blue band) and the $(w_0,w_a)$ parameterization (top-layered, green band). 
        The solid and dashed white lines indicate the best-fit reconstructions of $w(z)$, and the shaded regions are 68\% confidence-level uncertainties. 
        The horizontal dashed line denotes the $\Lambda$CDM prediction ($w=-1$). 
        Panels $(B_1$--$B_4)$ mirror these setups for the DESI DR2 BAO sample. 
        In all panels, the Big~Bang~Nucleosynthesis (BBN) prior and constraints on the CMB acoustic scale $\theta_*$ are imposed.
    }
    \label{fig:wz}
\end{figure*}

\begin{figure*}[htp]
    \centering
    \includegraphics[width=\textwidth]{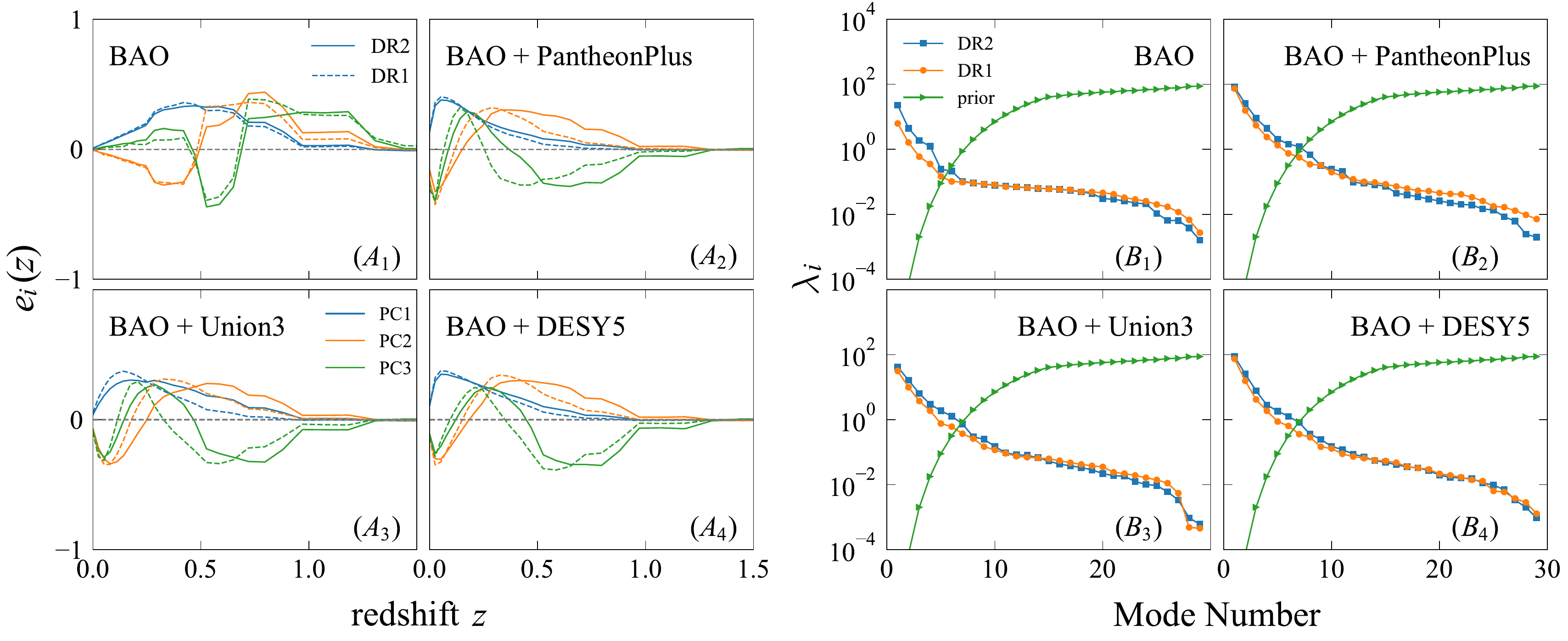}
    \caption{
{\bf Principal Component Analysis (PCA) of the dark energy equation of state $w(z)$. } The left panels $(A_1$--$A_4)$ plot the first three principal components $e_i(z)$ as functions of redshift, showcasing how different dataset combinations constrain distinct modes in $w(z)$. The right panels $(B_1$--$B_4)$ display the corresponding eigenvalues $\lambda_i$ versus the mode index, revealing the relative information that each dataset provides compared with the correlation-prior baseline (green curves). Solid lines distinguish DESI DR1 BAO analyses, and dashed lines indicate DESI DR2, and the Big~Bang~Nucleosynthesis (BBN) prior and constraints on the CMB acoustic scale $\theta_*$ are imposed in all cases. Larger eigenvalues signify better-constrained modes.}
    \label{fig:PCA}
\end{figure*}

The shape function $S_0(a)$ illustrates how dark energy grows relative to matter, and the function $S_1(a)$ reveals whether and how the dark energy pressure changes with time. Finally, $S_2(a)$—often referred to as the `state-finder' parameter \cite{SF1,SF2}—encodes both $w(a)$ and its derivative, making it particularly useful for distinguishing among dark energy models (See Methods for the defination and more details of the shape functions). For example, a `freezing' quintessence scenario predicts $S_2 < -1$ \cite{Caldwell:2005tm}, whereas barotropic fluid models require $S_2 > -1$ \cite{Scherrer:2005je}. If $S_2$ crosses the $-1$ threshold during cosmic evolution, it may indicate that multiple dark energy components dominate at different epochs.

Fig.~\ref{fig:shape_functions} displays the reconstructed $S$ functions derived from DESI BAO and three SNe datasets, based on Eqs.~\eqref{eq:chi_H}--\eqref{eq:DEfunctions}. All panels reveal a systematic deviation from the $\Lambda$CDM baseline, and this pattern of deviation is consistent across the four datasets, indicating a shared underlying trend. 

A key point is that the DESI~DR2 sample uncovers a stronger
dynamical DE signal than DESI~DR1. In particular, $S_0$ and $S_1$ both lie systematically above the $\Lambda$CDM expectation in DR2, pointing to a more pronounced departure from a pure cosmological constant. Even more notably, $S_2$ in DR2 exhibits a sharper crossing behavior relative to DR1, suggesting that the observed expansion may not be fully described by standard $\Lambda$CDM.

This amplified crossing, coupled with smaller uncertainties, bolsters the case for dark energy possessing additional internal degrees of freedom. Further reinforcing this conclusion, the DR2-based reconstructions closely match those obtained from three independent SNe samples. Their broad consistency indicates that these deviations probably reflect real dynamical evolution of the dark energy; for this not to be the case, a number of different systematics would all need to conspire in the same direction. Although DR2 provides stronger statistical evidence overall, uncertainties remain significant at $z \gtrsim 0.5$, limiting definitive conclusions at higher redshifts. Even so, the alignment between DESI~DR2 and SNe data underscores the importance of forthcoming high-precision measurements for clarifying the fundamental nature and potential evolution of dark energy.
\\

\section{Evolution of the equation of state}

To explore this further, we adopt the commonly used $w_0$--$w_a$ parameterization \citep{Chevallier2001,Linder2003}, in which the dark energy equation of state is written as $w(a) = w_0 + w_a\,(1 - a)$. Here, $w_0$ is the present-day equation-of-state value, while $w_a$ describes how $w$ evolves with redshift. Notably, $\Lambda$CDM is recovered by setting $w_0=-1$ and $w_a=0$. Fitting this model to the data constrains the possible dynamical nature of dark energy in a simple yet effective manner. Moreover, the $w_0$--$w_a$ approach serves as a bridge between the shape-function reconstructions and the fully non-parametric $w(z)$ methods discussed later, offering an intermediate step toward more general dark energy modeling.

\begin{figure*}[htp]
\includegraphics[width=0.9\textwidth]{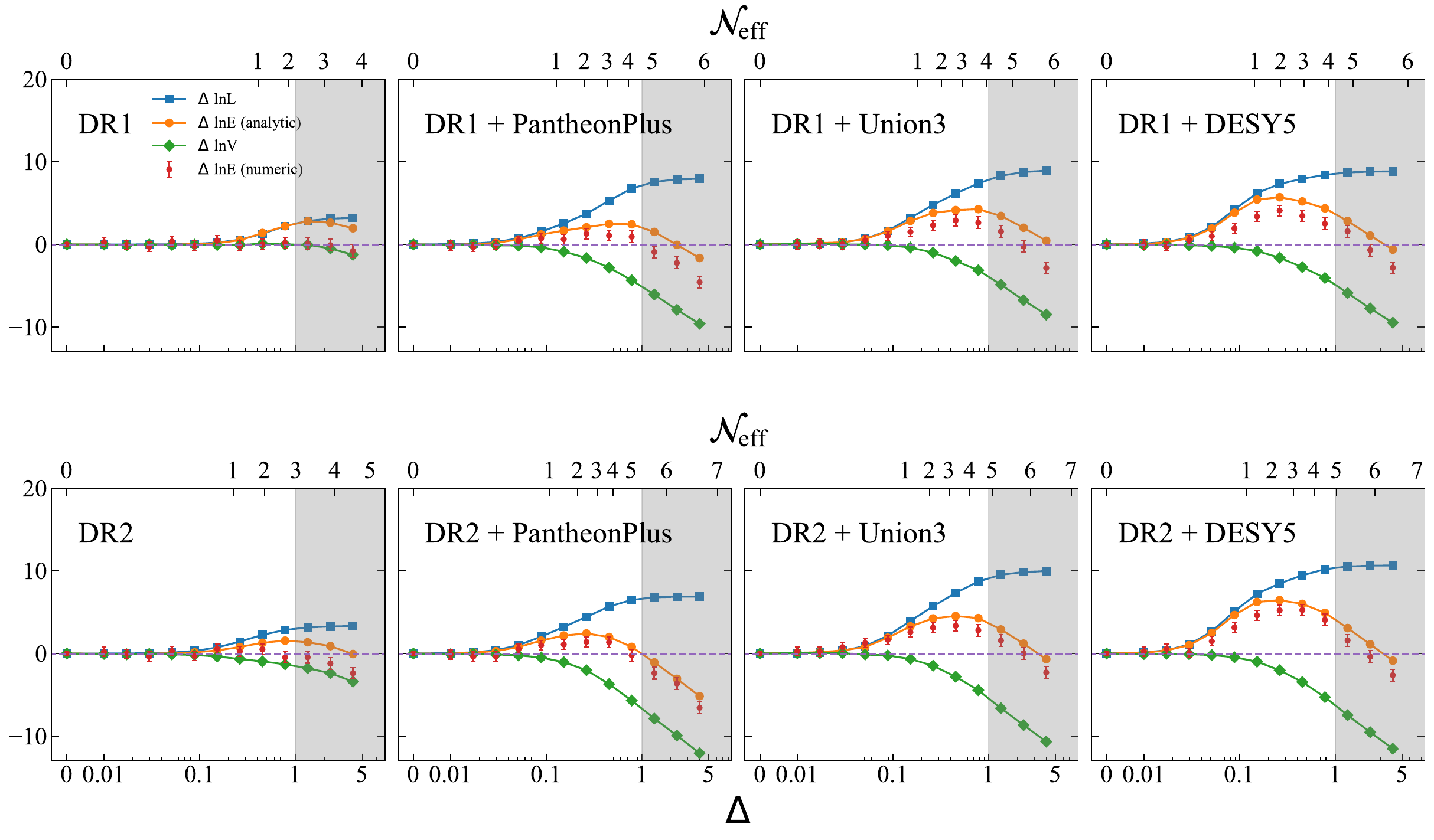}
\caption{
{\bf Likelihood, prior volume, and evidence as a function of the effective degrees of freedom \(\mathcal{N}_{\rm eff}\)}. Bayesian evidence peaks at ``moderate" support for dynamical $w(z)$ once SNe are included. The strength of this prior is varied by adding different values of $\Delta$ to its diagonal. For comparison, we show both the analytically estimated $\ln E$ (calculated as $\ln L + \ln V$) and the numerically computed values obtained via \texttt{Polychord}. The top panel presents results derived using DESI DR1 BAO data, while the bottom panel corresponds to DESI DR2. The Big~Bang~Nucleosynthesis (BBN) priors and constraints on the CMB acoustic scale $\theta_*$ are imposed in all cases. The shaded region highlights values of \(\Delta\) for which 
        the correlation prior is affected by the flat prior of the $w$ bins.
        See texts for further discussion.
}
\label{fig:evidence}
\end{figure*}

Fig.~\ref{fig:w0wa} shows the inferred constraints on $(w_0,\,w_a)$ for different data combinations (See Methods for datasets and priors used). The left panel corresponds to DESI DR1, while the right panel presents DR2. Contour colors indicate various datasets, with inner and outer contours marking 1$\sigma$ and 2$\sigma$ confidence intervals, respectively. We divide the parameter space into four dark energy models: Quintessence ($w>-1$ at all epochs, thus $w_0>-1$ and $w_0+w_a>-1$) \citep{quintessence}, Full Phantom ($w<-1$ at all epochs, thus $w_0<-1$ and $w_0+w_a<-1$) \citep{phantom}, Quintom~A ($w>-1$ in the past but $w<-1$ today, thus $w_0<-1$ and $w_0+w_a>-1$) \citep{quintom}, and Quintom~B ($w<-1$ in the past but $w>-1$ today, thus $w_0>-1$ and $w_0+w_a<-1$) \citep{quintom}.

Using only DESI~DR1 BAO measurements yields constraints that are broadly consistent with $\Lambda$CDM, albeit with mild ($\sim1.5$--$2\sigma$) hints of departure. Incorporating supernova data narrows these contours further, indicating a slight preference for dynamical dark energy. A similar pattern emerges from the DR2-only analysis, which achieves notably tighter constraints than DR1 and favors $w_0 > -1$ with $w_a < 0$, consistent with a Quintom~B-like scenario. Even so, $\Lambda$CDM remains viable at $\sim1.5\sigma$. This is weaker than the $2.3\sigma$ rejection of $\Lambda$CDM in \cite{DESI.DR2.BAO.cosmo} from DESI+CMB, because our CMB prior uses only part of the information, and also because we have chosen to inflate uncertainties. When DR2 is combined with supernova data, the best-fit values stay in the Quintom~B region, pushing the tension with $\Lambda$CDM above $2\sigma$,
showing somewhat stronger preference for departure from a simple cosmological constant
(see Supplementary Tables~1 and 2 for more details). 

To delve deeper into the dynamical nature of dark energy, we perform a non-parametric Bayesian reconstruction of $w(z)$ using the correlation-prior framework originally proposed in \cite{CPZ05,Fables11} and subsequently applied in \cite{Zhao2012,wrecon2017}. In this methodology, $w(z)$ is treated as a free function, while a correlation prior enforces smoothness and mitigates flat directions in the likelihood. The correlation prior itself can be motivated by theoretical considerations \cite{Raveri:2017qvt}; here, we adopt the version derived in \cite{Raveri:2017qvt} from Horndeski theory \cite{Horndeski:1974wa} (See Methods for details of the reconstruction).

Fig.~\ref{fig:wz} (white curves and blue bands) illustrates the best-fit and 68\%~CL constraints on $w(z)$ obtained from various data combinations. The upper row corresponds to DESI DR1 BAO data, and the lower row to DR2. In each row, the leftmost panel reports BAO-only constraints, while the other panels incorporate additional supernova datasets: PantheonPlus, Union3, and DESY5. In all cases, $w(z)\neq -1$ is clearly indicated. Throughout the paper, we define the significance (i.e., the signal-to-noise ratio, SNR) of the reconstructed $w(z)$ deviating from a specific model $w_{\rm mod}$ via
\begin{equation}
\mathrm{SNR}^2 \equiv (\mathbf{w}-\mathbf{w_{\rm mod}})^T\,\mathbf{C}_{\rm w}^{-1}\,(\mathbf{w}-\mathbf{w_{\rm mod}}),
\end{equation} where $\mathbf{C}_{\rm w}$ is the posterior covariance for the $w$ bins. For the DR1-based reconstructions (panels $A_1$--$A_4$), the SNR of $w\neq -1$ is $2.6$, $3.9$, $3.9$, and $4.2$, respectively. Switching to the DR2 dataset yields comparable significance at $2.6$, $3.7$, $4.3$, and $4.5$, respectively. It is interesting that the preference for deviations from $\Lambda$CDM when the SNe samples are included remained strong with the enhanced statistical power of DESI DR2.

All dataset combinations exhibit a persistent pattern in $w(z)$: $w>-1$ at $z\lesssim 0.2$ and $w<-1$ at $z\sim0.75$, which is consistent with the results reported in companion DESI papers \citep{DESI.DR2.BAO.cosmo,Y3.cpe-s1.Lodha.2025}. Including supernova data highlights this feature further, revealing pronounced oscillations in $w(z)$. These patterns remain stable and gain additional significance when DR1 BAO is replaced by DR2. In fact, oscillatory or non-monotonic $w(z)$ evolution arises naturally in several classes of dark energy models – including axion-like quintessence scenarios with periodic potentials \citep{Frieman:1995pm}, interacting dark sector frameworks, where energy exchange between dark energy and dark matter can induce transient oscillations in $w$ \citep{Amendola:1999er,Wang:2016lxa}, and multi-field quintom constructions that allow $w(z)$ to cross the $-1$ threshold and back \citep{quintom,Feng:2004ff}. Crucially, the oscillatory feature seen in our reconstruction is not an artifact of any single dataset: it appears consistently across different Type Ia supernova samples  and remains essentially unchanged when the BAO dataset is upgraded from DESI DR1 to DR2. This stability across independent data sets and updated observations reinforces the interpretation that the observed oscillations in $w(z)$ are a robust physical signal, potentially pointing toward an underlying dynamical dark energy component consistent with the aforementioned model classes.

The green bands—reconstructions using the CPL parameterization—track the transition across $w=-1$ well and are broadly compatible with the non-parametric results, reinforcing the robustness of these conclusions.

While the significance of $w \neq -1$ stems from the reduction in $\chi^2$ relative to $\Lambda$CDM, fully assessing the fit also requires determining the effective degrees of freedom (DoF) in the reconstructed $w(z)$. This step is nontrivial because the correlation prior couples the $w(z)$ bins. To address this, we perform a principal component analysis (PCA) \cite{Huterer:2002hy} to decorrelate the prior and posterior inverse covariance matrices of $w(z)$. Further technical details on the PCA procedure can be found in the Supplementary Information.

Fig.~\ref{fig:PCA} shows the PCA results for the reconstructed dark energy equation of state \( w(z) \). The left panels \((A_1\text{--}A_4)\) plot the eigenvectors of the first few well-constrained modes, highlighting their redshift dependence, while the right panels \((B_1\text{--}B_4)\) display the corresponding eigenvalues \(\lambda_i\) for different dataset combinations. The PCA identifies independent modes constrained by the data. In panel \(A_1\) (BAO-only), the first principal component (PC1) is smooth and relatively featureless, indicating that BAO data alone primarily constrain a single dominant mode of \( w(z) \). By contrast, adding supernova data (panels \(A_2\text{--}A_4\)) produces additional constrained modes; PC2 and PC3 become more oscillatory, especially at low redshift, demonstrating that supernova observations significantly improve constraints on the redshift evolution of \( w(z) \). Because the main feature of $w(z)$ is effectively a linear combination of these three modes, further progress will come from shrinking the PC1 error bar through improved low-$z$ SN calibration, tightening PC2 and PC3 with denser distance tracers in the $0.3<z<0.8$ window, and extending observations to $z\!\gtrsim\!0.8$ to assess whether additional PCs are required.

The right panels show the eigenvalues \(\lambda_i\), which quantify the statistical significance of each mode (larger values correspond to tighter constraints). In panel \(B_1\) (also BAO-only), four eigenvalue notably exceeds that of the correlation prior (green curve). 
When supernova datasets are incorporated (panels \(B_2\text{--}B_4\)), more modes become well constrained. Comparing DESI Data Release~1 (DR1; orange) and Data Release~2 (DR2; blue) eigenvalues reveals that DR2 consistently achieves larger eigenvalues for the constrained modes, indicating stronger statistical power in the DR2 sample. This gain is most evident when supernova data are included, as the gap between the first few eigenvalues of DR1 and DR2 widens, showing that DR2 improves constraints across multiple modes of \( w(z) \). We further validate these findings by computing the effective number of degrees of freedom $\mathcal{N}_{\rm eff}$ numerically with covariant principal component analysis (CPCA) \cite{Raveri:2021dbu}, detailed in the Supplementary Information. The CPCA results align with the PCA, confirming the robustness of our assessment of how many independent modes of \( w(z) \) are constrained by the data.

We can project $w(z)$ onto the PC eigenmodes $1+w(z) = \sum_i \gamma_i e_i(z)$ where the corresponding eigenvalue $\lambda_i$ of the covariance matrix represents the error of the amplitude $\gamma_i$, $\lambda_i =1/\sigma^2(\gamma_i)$. Supplementary Table 5 shows the amplitude of the first five PCs. 
Considering the case with DR2 and DR1/DR2 + SNe, 
the amplitude of the first PC is positive, indicating that this mode describes a behavior of $w(z)$ that increases from $-1$ over time. 
The improvement in the fits, $\Delta \chi^2 = - \sum_i [\gamma_i/\sigma(\gamma_i)]^2$, is dominated by PC1. 
Including higher-order PCs, multiple crossings of $w=-1$ occur, as we observe in the reconstructed $w(z)$ shown in Fig.~\ref{fig:wz}. In the case of DR2 + SNe, PC4 gives the second largest contribution to $\Delta \chi^2 $.

It is the advantage of the Bayesian reconstruction that the dependence of the reconstructed $w(z)$ on the correlation prior is explicit. To demonstrate this, we deliberately change the amplitude of the correlation prior, $A$, and impose a stronger prior to avoid possible overfitting. By tuning $A$, the effective number of degrees of freedom $\mathcal{N}_{\rm eff}$ can be reduced. In the case of $\mathcal{N}_{\rm eff}=2$, the reconstructed $w(z)$ is consistent with the results that we obtained based on the $w_0$-$w_a$ parametrization as shown in Supplementary Fig. 1, with related $\Delta\chi^2$ shown in Supplementary Table 6.
\\

\section{Bayesian evidence and evolution}

We evaluate the evidence against \(\Lambda\)CDM analytically and numerically (see Methods for details), as illustrated in Fig.~\ref{fig:evidence}. Consistent with expectations, \(\ln L\) grows while \(\ln V\) decreases monotonically with increasing \(\Delta\) (or \(N_{\rm eff}\)), which reflects the number of modes in \( w(z) \) that remain after applying the correlation prior. The analytic and numerical evidence estimates typically agree, except in BAO-only scenarios, where the data are not sufficiently constraining to maintain Gaussianity in the posterior. Fig.~\ref{fig:evidence} plots the model evidence and related quantities versus \(\Delta\), with the effective degrees of freedom \(N_{\rm eff}\) noted on the top axis. The top panels show DESI DR1 results, and the bottom panels DESI DR2. We display the change in log-likelihood \(\Delta\ln L\) (blue squares), the analytic \(\Delta\ln E\) (orange circles), the numerical \(\Delta\ln E\) (red points), and the change in log-prior volume \(\Delta\ln V\) (green diamonds).

The increase in both $\Delta \ln E$ and the signal-to-noise ratio (SNR) for detecting $w \neq -1$ with larger $\mathcal{N}_{\rm eff}$ further supports the idea that adding extra degrees of freedom favors a dynamical dark energy interpretation over a pure cosmological constant. When using BAO data alone, $\Delta \ln E$ remains consistent with zero for all $\mathcal{N}_{\rm eff}$, confirming that BAO measurements by themselves offer limited evidence against $\Lambda$CDM. In contrast, once supernova datasets are added, $\Delta \ln E$ rises substantially. For instance, DR1 + DESY5 yields a peak $\Delta \ln E$ of $4.1 \pm 0.6$ (classified as “Moderate” on Jeffreys’ scale \cite{Jeffreys1939,Trotta:2008qt,LeeWagenmakers2014}) at $N_{\rm eff}=2$, increasing to $5.2 \pm 0.7$ (also “Moderate”) for DR2 + DESY5 at $N_{\rm eff}=3$. A similar pattern emerges with BAO + Union3: the evidence systematically increases from DR1 to DR2, reaching $3.3 \pm 0.7$ (“Moderate”) at $N_{\rm eff}=3$ for DR2.

By contrast, BAO + PantheonPlus shows weaker evidence, changing only slightly from DR1 to DR2, with $\Delta \ln E$ peaking at $1.5 \pm 0.6$ (“Weak”) in DR1 and $1.4 \pm 0.7$ (“Weak”) in DR2—consistent with what is seen under the CPL parameterization. Although the PantheonPlus dataset alone can show a mild preference for $w\neq -1$ (see Supplementary Fig.~2), combining it with BAO data partially cancels that signal. One possibility is a mild tension between the PantheonPlus-inferred distance scale and the BAO-preferred expansion history, pushing the joint constraints closer to $\Lambda$CDM. Further investigation is needed to determine whether this is attributable to systematics.

To gauge the strongest possible deviation from $w = -1$ without overfitting the data, we look at $w(z)$CDM models with the largest departure from $\Lambda$CDM while still having a positive evidence, {\i.e.} $\Delta \ln E \geq 0$. Under this condition, the significance of $w \neq -1$ for BAO + DESY5 reaches $4.1$ (DR1) and $4.3$ (DR2). For BAO + Union3, the corresponding values are $3.8$ (DR1) and $3.9$ (DR2). Meanwhile, BAO + PantheonPlus yields maximum SNRs of $3.2$ (DR1) and $3.1$ (DR2), reinforcing the prior finding that PantheonPlus offers comparatively weaker support for $w \neq -1$ than either Union3 or DESY5.
\\

\section{Summary and conclusions}

With the start of the DESI survey, we now have access to exceptionally precise BAO measurements, providing a robust tool to probe the nature of dark energy. In this work, we analyzed DESI DR2 BAO observations in tandem with complementary cosmic distance indicators— SNe data and the CMB acoustic scale \(\theta_*\)—to test for deviations from the \(\Lambda\)CDM paradigm. Employing a shape-function approach \cite{Gu:2024jhl,Wang:2024qan}, we derived key observables sensitive to dark energy (DE) and found consistent evidence for an evolving equation of state \( w(z) \) across multiple, independent datasets.

Building on these findings, we performed a non-parametric Bayesian reconstruction of \( w(z) \) by jointly analyzing DESI BAO, SNe, and CMB data, employing a correlation prior rooted in Horndeski theory. Our results exhibit strong statistical support for dynamical dark energy, at a confidence level exceeding \( 3\sigma \). Rigorous model selection further indicates that DESI BAO, SNe, and CMB observations together constrain roughly three independent degrees of freedom (DoFs) without overfitting. In particular, combining DESI BAO, CMB, and the five-year DES supernova sample detects dark energy evolution at a signal-to-noise ratio of $3.9$, with a Bayesian evidence of \( 5.2 \pm 0.7 \) in favor of \( w(z) \)CDM over \(\Lambda\)CDM when \( N_{\rm eff} = 3 \).

These results suggest that dark energy may not be described by a simple cosmological constant, favoring a time-dependent equation of state \( w(z) \) instead. Although earlier work \cite{wrecon2017} hinted at similar conclusions, this study provides some of the tightest constraints to date, thanks in large part to the enhanced statistical power of DESI DR2 BAO data. Furthermore, adding supernova datasets significantly boosts sensitivity to potential dark energy evolution: all three independent SN samples (PantheonPlus, Union3, and DESY5) show broadly consistent indications of deviations from \(\Lambda\)CDM. 

Companion DESI papers \cite{DESI.DR2.BAO.cosmo,Y3.cpe-s1.Lodha.2025} investigate the time evolution of dark energy using complementary methods, and also report the detection of the dynamical dark energy. The level of significance is generally greater in those works because the full Planck measurements are used there, as opposed to the $\theta_*$ information used in this work. Moreover, the recent Atacama Cosmology Telescope (ACT) results find that the time evolution of dark energy is still favored when replacing the Planck data with the ACT measurements \cite{ACT:2025tim}.

Extensive cross‑checks indicate that known systematics cannot mimic the oscillatory pattern in Fig. \ref{fig:wz}. Photometric offsets and a possible evolution of the colour–luminosity relation in PantheonPlus can shift $w(z)$ by about $0.03$ at $z\lesssim0.4$ \citep{Brout:2022}, and the $0.04$ mag low–high‑$z$ mismatch removes only a smooth trend in $(w_0,w_a)$ fits when using the DESY5 SNe sample \citep{Efstathiou2025}. For the BAO analysis, DESI DR1 analyses bound template‑ and reconstruction‑induced BAO shifts to 0.1\% \citep{DESI2024.III.KP4}. All of these effects vary smoothly with redshift: they can tilt or raise the whole $w(z)$ curve but cannot produce the alternating‑sign wiggles we recover. The feature therefore remains most naturally interpreted as evidence for genuine dark‑energy dynamics.

In conclusion, our study finds tentative evidence for departures from \(\Lambda\)CDM using the DESI DR2 data. 
While our results favor a dynamical dark energy interpretation, continued observational and theoretical efforts will be critical in establishing whether these deviations signify new physics or stem from unknown systematics. The complete DESI sample, with increased statistical depth and extended redshift coverage, is expected to yield even stricter constraints on dark energy evolution. Additional surveys, such as Euclid \cite{Amendola:2016saw} and the Prime Focus Spectrograph (PFS) \cite{PFSTeam:2012fqu}, will deliver complementary BAO and large-scale structure measurements, enabling cross-validation of DESI results. Likewise, next-generation CMB projects will improve knowledge of early-universe physics, strengthening the overall dark energy picture.

\section{Methods}

\subsection{Datasets and priors}

DESI at Kitt Peak National Observatory in Arizona marks a significant step forward in Stage~IV large-scale structure surveys \citep{DESI2016a.Science,DESI2022.KP1.Instr}. Designed with a $3.2^\circ$ diameter prime focus corrector \citep{Corrector.Miller.2023}, DESI deploys 5{,}000 fibers \citep{FiberSystem.Poppett.2024} via a robotic focal plane assembly \citep{FocalPlane.Silber.2023} to simultaneously measure multiple spectra. These hardware capabilities are bolstered by a high-throughput spectroscopic data processing pipeline \citep{Spectro.Pipeline.Guy.2023} and an efficient operations plan \citep{SurveyOps.Schlafly.2023}. Since beginning operations in 2021, DESI has been gathering high-fidelity spectra for numerous target classes---Bright Galaxy Sample (BGS; $0.1 < z < 0.4$) \citep{BGS.TS.Hahn.2023}, Luminous Red Galaxies (LRGs; $0.4 < z < 1.1$) \citep{LRG.TS.Zhou.2023}, Emission Line Galaxies (ELGs; $1.1 < z < 1.6$) \citep{ELG.TS.Raichoor.2023}, Quasars (QSOs; $0.8 < z < 2.1$) \citep{QSO.TS.Chaussidon.2023}, and the Lyman-$\alpha$ Forest ($1.77 < z < 4.16$) \citep{KP6s4-Bault}. Early data analyses and survey validation \citep{DESI2022.KP1.Instr,DESI2023b.KP1.EDR} confirmed that DESI satisfies the performance benchmarks required for a Stage~IV survey, advancing our grasp of dark energy's role in cosmic expansion \citep{DESI2016a.Science}.

DESI's DR1 spans observations from May~14, 2021, to June~14, 2022 \citep{DESI2024.I.DR1}, and has already yielded new insights into dark energy by detecting the baryon acoustic oscillation (BAO) signature in galaxy and quasar clustering \citep{DESI2024.III.KP4}, as well as in the Lyman-$\alpha$ forest \citep{DESI2024.IV.KP6}. These results fit seamlessly with external data \citep{DESI2024.VI.KP7A}, paving the way for more comprehensive analyses that incorporate the full clustering information from the DESI galaxy sample and other tracers \citep{DESI2024.VII.KP7B,DESI2024.V.KP5}. The subsequent DR2, covering observations through April~9, 2024 \citep{DESI.DR2.BAO.lya,DESI.DR2.BAO.cosmo}, not only encompasses the entirety of DR1 but also extends the redshift range and statistical reach of the survey. This expanded dataset offers a critical test of dynamical dark energy: if dark energy truly evolves over cosmic time, the enriched DR2 sample could reveal a more pronounced signature. By comparing DR1 and DR2 results, one can assess both internal consistency and potential evidence for evolving dark energy. This paper contributes to a suite of companion analyses that build on the main cosmological results \cite{DESI.DR2.BAO.cosmo}, with related supporting study on dark energy \cite{Y3.cpe-s1.Lodha.2025} and neutrino constraints \cite{Y3.cpe-s2.Elbers.2025}.

Our analysis draws upon three categories of measurements: BAO observations from DESI DR1 and DR2 \cite{DESI2024.III.KP4,DESI2024.IV.KP6}, SNe samples from PantheonPlus, Union3, and the 5-year DES survey (DESY$5$), and distance constraints from the CMB. Because dark energy chiefly affects the Universe's background expansion history, and modeling its perturbations requires additional assumptions, we focus here on background observables. Specifically, we incorporate the DESI DR1/DR2 BAO measurements, SNe luminosity distances, and the Big~Bang~Nucleosynthesis (BBN) prior $\Omega_{\mathrm{b}}h^2=0.02196\pm0.00063$ \cite{BBN}. We also include CMB distance information from Planck PR4 \cite{Rosenberg:2022sdy}, imposing $100\,\theta_*=1.04098\pm0.00042$, where $\theta_*\equiv r_*/D_{\mathrm{M}}(z_*)$ is the ratio of the comoving sound horizon at recombination $r_*$ to the transverse comoving distance $D_{\mathrm{M}}(z_*)$ \cite{Planck18,DESI2024.VI.KP7A}. The quantity $r_*$ is computed using \texttt{RECFAST} \cite{Recfast} within \texttt{CAMB} \cite{CAMB}. As in \cite{DESI2024.VI.KP7A}, we inflate the error bar from \cite{Rosenberg:2022sdy} (column `PR4\_12.6' TTTEEE in Table~5) by $75\%$ to account for potential broadening of $\theta_*$ constraints in extended cosmological models.

Our baseline dataset comprises DESI BAO measurements, the BBN prior on $\Omega_{\mathrm{b}}h^2$, and the CMB acoustic-scale constraint on $\theta_*$. We then supplement this baseline with each of the three SNe samples in turn: `DR1(2)+PantheonPlus', `DR1(2)+Union3', and `DR1(2)+DESY5' for the DES 5-year sample. Each combination is analyzed independently to gauge its impact on $w(z)$ constraints.

\subsection{The shape functions of dark energy}

Galaxy surveys measure cosmic distances via $(D_{\rm M}/r_{\rm d},\,D_{\rm H}/r_{\rm d})$ 
or $D_{\rm V}/r_{\rm d}$ at specific redshifts, where $D_{\rm H}\equiv c/H$ with $H$ being the Hubble expansion rate, and $D_{\rm M}$ the comoving distance. The volume-averaged distance is defined by $D_{\rm V}\equiv \bigl(z\,D_{\rm M}^2 D_{\rm H}\bigr)^{1/3}$, with $z$ being the redshift, and $r_{\rm d}$ the sound horizon scale. These observations constrain the shape of $H(z)$, which is directly related to dark energy. To link $D_{\rm H}$ and $D_{\rm M}$, we adopt the parametrization introduced in \cite{Zhu2014}, providing an accurate yet general description of cosmic distances across a wide range of cosmologies:
\begin{eqnarray}
\label{eq:chi_H}
\frac{D_{\rm M}(z)}{D_{\rm M}^{\rm f}(z)}\,R &=& \alpha_0\left(1 + \alpha_1 x + \alpha_2 x^2 + \alpha_3 x^3 + \dots\right),\nonumber\\
\frac{D_{\rm H}(z)}{D_{\rm H}^{\rm f}(z)}\,R &=& \beta_0\left(1 + \beta_1 x + \beta_2 x^2 + \beta_3 x^3 + \dots\right),
\end{eqnarray}
where the superscript (subscript) `f' denotes quantities evaluated in a fiducial \(\Lambda\)CDM model with $\Omega_{\rm M}=0.3153$.
The normalization factor $R \equiv r_{\rm d}^{\rm f}/r_{\rm d}$ accounts for discrepancies in the sound horizon scale, and $x \equiv D_{\rm M}^{\rm f}(z)/D_{\rm M}^{\rm f}(z_{\rm p}) - 1$, where $D_{\rm M}$ denotes the comoving distance and $z_{\rm p}$ is a pivot redshift for the series expansion.

Because $D_{\rm H}$ and $D_{\rm M}$ are interdependent, the coefficients $\beta_i$ follow from $\alpha_i$ via
\begin{eqnarray}
\label{eq:alpha_beta}
\beta_0 = \alpha_0\bigl(1+\alpha_1\bigr), 
\quad
\beta_i = (i+1)\,\frac{\alpha_i+\alpha_{i+1}}{1+\alpha_1}
\quad
(i>0).
\end{eqnarray}
To avoid overfitting, the series expansion is truncated once the uncertainty in any $\alpha_i$ exceeds unity. For DESI~DR1 and DR2 BAO data, a cutoff at $i=3$ (yielding four $\alpha$ parameters) proves adequate.

Given the values of $\alpha$, one can construct the following function \citep{Gu:2024jhl,Wang:2024qan}:
\begin{equation}
S(a) \;\equiv\; A\,H^2(a)\,a^3 \;=\; B\,X(a)\,a^3\;+\;C,
\end{equation}
where $A,B,$ and $C$ are constants, and $X(a)\equiv \rho_{\rm DE}(a)/\rho_{\rm DE}(1)$ is the normalized dark energy density. Since $S(a)$ shares the same functional form as $\rho_{\rm DE}(a)$—up to a shift and overall rescaling—this `shape function' effectively encodes the potential dynamical properties of dark energy. 

Specifically, we define three dimensionless functions that offer clear diagnostics for dark energy evolution:
\begin{eqnarray}
\label{eq:DEfunctions}
S_0(a)&\equiv&a^3-\frac{3\left[S(a)-S(1)\right]}{S'(1)}
\;=\;a^3+\frac{X(a)\,a^3-1}{w(1)}
\xrightarrow{\Lambda}1, 
\quad\nonumber\\
S_1(a)&\equiv&
\frac{1}{a^3}\,\frac{S'(a)}{S'(1)}
\;=\;\frac{P_{\rm DE}(a)}{P_{\rm DE}(1)}
\xrightarrow{\Lambda}1, 
\quad\nonumber\\
S_2(a)&\equiv&-\frac{S''(a)}{3\,S'(a)}
\;=\;w(a)\;-\;\frac{w'(a)}{3\,w(a)}
\xrightarrow{\Lambda}-1,
\end{eqnarray}
where derivatives are taken with respect to $\ln a$. Here, $P_{\rm DE}(a)$ and $w(a)$ denote the pressure and equation of state of dark energy, respectively, and the arrows indicate their expected values under $\Lambda$CDM.

\subsection{A non-parametric Bayesian reconstruction of dark energy with the Horndeski prior}

Within the $w(z)$ cold dark matter (CDM) framework, $w(z)$ is discretized into $N=29$ piecewise-constant segments, $\mathbf{w}\equiv\{w_1,w_2,\dots,w_N\}$, uniformly spaced in scale factor for $z<2.5$. Additionally, an extra `wide' bin $w_{\mathrm{wide}}=-1$ is assigned for $z>2.5$ up to recombination. Allowing $w_{\mathrm{wide}}$ to vary left it largely unconstrained by our data, and it showed minimal correlations with the other $w$ bins. We therefore fix $w_{\mathrm{wide}}=-1$ for simplicity. Alongside these $w_i$, we vary several cosmological parameters: $\Omega_{\rm M}$, the current fractional matter density; $\Omega_{\rm b}h^2$, the physical baryon density; $H_0$, the Hubble constant; and $\mathcal{M}_{\mathrm{b}}$, the absolute SNe magnitude (used only when supernova data are included).

We place sufficiently broad flat priors on all parameters. Specifically, 
$H_0 \in [20,100]$, 
$\Omega_{\rm M} \in [0.01,0.99]$, 
$w_0 \in [-3,1]$, 
$w_a \in [-3,2]$, 
and $w_i \in [-6,4]$ (for $i=1,\dots,29$).
We explore the resulting parameter space using the Markov Chain Monte Carlo (MCMC) algorithm implemented in {\tt Cobaya} \cite{Torrado:2021}. At each step, we compute a total $\chi^2$ that combines contributions from both the data and the correlation prior. 
The prior covariance $\mathbf{C}_{\Pi}$ is derived from Horndeski-based simulations \cite{Raveri:2017qvt} spanning a broad parameter space, as investigated in \cite{Raveri:2017qvt}. Specifically, the prior is defined as,
\begin{eqnarray}
    C(a,a')&\equiv&\langle[w(a)-w_{\rm fid}(a)][w(a)-w_{\rm fid}(a')]\rangle \nonumber\\
    &=&\sqrt{C(a)C(a')}R(a,a'), 
\end{eqnarray} where $C(a)\equiv C(a,a)$ and $R(a,a')$ is the normalised correlation function thus equals unity for $a=a'$. The functional form of the correlation prior used in this work is taken from \cite{Raveri:2017qvt}:
\begin{eqnarray}
C(a)&=&0.05+0.8 a^2, \nonumber\\
R(a,a')&=&{\rm exp}\left[-\left(\left|{\rm ln} \ a-{\rm ln} \ a'\right|/0.3\right)^{1.2}\right].
\end{eqnarray}

The total $\chi^2$ minimised in the MCMC process has two pieces, namely\begin{eqnarray}
\chi^2&=&\chi_{\rm data}^2+A\chi_{\rm prior}^2,  \nonumber\\
\chi_{\rm prior}^2&=&{\bf D}_{\rm w}^T {\bf C}^{-1}_{\Pi}{\bf D}_{\rm w} 
=\mathbf{w}^T \tilde{\mathbf{C}}^{-1}_{\Pi} \mathbf{w},
\end{eqnarray} where ${\bf D}_{\rm w}={\bf w}-{\bf w}_{\rm fid} =\left({\bf I}-{\bf S}\right){\bf w}$ and $\tilde{\mathbf{C}}^{-1}_{\Pi}$, the inverse modified covariance matrix of the prior, is defined as $\tilde{\mathbf{C}}^{-1}_{\Pi}\equiv \left({\bf I}-{\bf S}\right)^T \mathbf{C}^{-1}_{\Pi} \left({\bf I}-{\bf S}\right)$. Matrices ${\bf I}$ and ${\bf S}$ are the identity matrix and the transformation matrix, respectively, and the fiducial model ${\bf w}_{\rm fid}={\bf Sw}$ is calculated by taking a local average of 5 neighboring $w$ bins through the transformation matrix $S$ \cite{Fables11,Zhao2012,wrecon2017,Xrecon2018}. The factor $A$ balances the strength between data and prior, and we set $A=1$ as a default to obtain the main result in Fig. \ref{fig:wz}. 

We validate our pipeline using mock datasets generated for four input dark energy models with the same data covariance matrix for DESI and SNe measurements, before applying to actual observations. Further technical details are provided in the Supplementary Information.

\subsection{The calculation of the Bayesian evidence}

For a rigorous test of whether the \( w(z) \)CDM model is preferred over \(\Lambda\)CDM, we evaluate the Bayesian evidence using both {\tt PolyChord} \cite{Handley:2015fda,Handley:2015vkr} within {\tt Cobaya} \cite{Torrado:2021} and an analytic approximation. Under the assumption that both prior and posterior distributions are Gaussian, the evidence \(E\) can be computed as \cite{Zhao2012} 
\({\rm ln}E= {\rm ln}V+{\rm ln}L,\)
where 
\({\rm ln}V=\tfrac{1}{2}\bigl({\rm ln\,det}\,\mathbf{C}_{\rm post}-{\rm ln\,det}\,\tilde{\mathbf{C}}_{\rm prior}\bigr)\).
Here, \(\tilde{\mathbf{C}}_{\rm prior}\) and \(\mathbf{C}_{\rm post}\) are the modified correlation-prior covariance and the posterior covariance, respectively, and \({\rm ln}L\) is the maximum log-likelihood from the data and prior. Because our reconstruction uses a running-average method to specify the fiducial \( w(z) \), \(\tilde{\mathbf{C}}_{\rm prior}\) has zero eigenvalues that render its volume ill-defined \cite{Fables11,Zhao2012} (see Supplementary Information).

To address this, we add a diagonal term \(\Delta^{-2}\) to the prior correlation matrix, interpolating between \(\Lambda\)CDM (\(\Delta=0\)) and a \( w(z) \)CDM model in which \(\mathbf{w}_{\rm fid}\) is unconstrained by the correlation prior. The prior volume is controlled by $\Delta$ and, for $\Delta <1$, it becomes independent of the flat prior imposed on $\mathbf{w}$ in the MCMC analysis. The parameter $\Delta$ changes the effective number of degrees of freedom, $\mathcal{N}_{\rm eff}$, by controling the strength of the correlation prior.

\vspace{1cm} 

\begin{center}
{\bf Data availability statement}
\end{center}

The data used in this analysis will be made public along the Data Release 2 (details in \url{https://data.desi.lbl.gov/doc/releases/}).
The data files and python scripts used to produce figures in the main text of this paper are available
in a Zenodo repository \cite{Zenodo}.

\acknowledgments 

We thank Dragan Huterer, Rodrigo Calderon, Eric Linder and Paul Martini for discussions. GG, XW, YW, RZ and GBZ are supported by NSFC grant 12525301, and by the CAS Project for Young Scientists in Basic Research (No. YSBR-092). GBZ is also supported by the New Cornerstone Science Foundation through the XPLORER prize. YW is supported by National Key R\&D Program of China No. (2023YFA1607800, 2023YFA1607803, 2022YFF0503404), NSFC Grants (12273048, 12422301), and the Youth Innovation Promotion Association CAS. JLCC is supported by the CONAHCYT grant CBF2023-2024-589. 
KK is supported by STFC grant
number ST/W001225/1. 

This material is based upon work supported by the U.S. Department of Energy (DOE), Office of Science, Office of High-Energy Physics, under Contract No. DE–AC02–05CH11231, and by the National Energy Research Scientific Computing Center, a DOE Office of Science User Facility under the same contract. Additional support for DESI was provided by the U.S. National Science Foundation (NSF), Division of Astronomical Sciences under Contract No. AST-0950945 to the NSF’s National Optical-Infrared Astronomy Research Laboratory; the Science and Technology Facilities Council of the United Kingdom; the Gordon and Betty Moore Foundation; the Heising-Simons Foundation; the French Alternative Energies and Atomic Energy Commission (CEA); the National Council of Humanities, Science and Technology of Mexico (CONAHCYT); the Ministry of Science, Innovation and Universities of Spain (MICIU/AEI/10.13039/501100011033), and by the DESI Member Institutions: \url{https://www.desi.lbl.gov/collaborating-institutions}.

The DESI Legacy Imaging Surveys consist of three individual and complementary projects: the Dark Energy Camera Legacy Survey (DECaLS), the Beijing-Arizona Sky Survey (BASS), and the Mayall z-band Legacy Survey (MzLS). DECaLS, BASS and MzLS together include data obtained, respectively, at the Blanco telescope, Cerro Tololo Inter-American Observatory, NSF’s NOIRLab; the Bok telescope, Steward Observatory, University of Arizona; and the Mayall telescope, Kitt Peak National Observatory, NOIRLab. NOIRLab is operated by the Association of Universities for Research in Astronomy (AURA) under a cooperative agreement with the National Science Foundation. Pipeline processing and analyses of the data were supported by NOIRLab and the Lawrence Berkeley National Laboratory. Legacy Surveys also uses data products from the Near-Earth Object Wide-field Infrared Survey Explorer (NEOWISE), a project of the Jet Propulsion Laboratory/California Institute of Technology, funded by the National Aeronautics and Space Administration. Legacy Surveys was supported by: the Director, Office of Science, Office of High Energy Physics of the U.S. Department of Energy; the National Energy Research Scientific Computing Center, a DOE Office of Science User Facility; the U.S. National Science Foundation, Division of Astronomical Sciences; the National Astronomical Observatories of China, the Chinese Academy of Sciences and the Chinese National Natural Science Foundation. LBNL is managed by the Regents of the University of California under contract to the U.S. Department of Energy. The complete acknowledgments can be found at \url{https://www.legacysurvey.org/}.

Any opinions, findings, and conclusions or recommendations expressed in this material are those of the author(s) and do not necessarily reflect the views of the U. S. National Science Foundation, the U. S. Department of Energy, or any of the listed funding agencies.

The authors are honored to be permitted to conduct scientific research on I'oligam Du'ag (Kitt Peak), a mountain with particular significance to the Tohono O’odham Nation.

\begin{center}
{\bf Author contributions}
\end{center}

G.G. and X.W. carried out the data analysis and generated all figures under the supervision of G.B.Z. and Y.W. Y.W. contributed to the Bayesian-evidence calculation and assisted with manuscript preparation. G.B.Z. conceived the project, designed the analysis strategy, and led the writing of the paper.
L.P. proposed the use of a Horndeski correlation prior and provided technical support for its implementation. L.P. and K.K. offered critical guidance on the presentation and interpretation of the results, and made important contribution to the manuscript. J.A.P., Z.C., J.L.C., M.I., A.S., R.Z. and J.P.K. contributed valuable comments and edits. All remaining co-authors are core DESI “builder’’ members who were responsible for survey design, instrument construction, data acquisition, and data processing. 

\begin{center}
{\bf Competing Interests}
\end{center}

The authors declare that they have no competing interests.

\clearpage


\bibliographystyle{naturemag}
\bibliography{references}


\clearpage
\onecolumngrid

\begin{center}
\begin{Large}
{\bf Supplementary Information}
\end{Large}
\end{center}

This document includes Supplementary Figures 1-7, Supplementary Discussions, and Supplementary Tables 1-6.


\setcounter{figure}{0}
\setcounter{table}{0}
\renewcommand{\figurename}{{\bf Supplementary Fig.}}
\renewcommand{\tablename}{{\bf Supplementary Table}}
\renewcommand{\thetable}{\arabic{table}}

\setcounter{section}{0}
\setcounter{secnumdepth}{1}
\renewcommand\thesection{\arabic{section}}

\section{Reconstruction of $w(z)$ using a Wiener filter projection}

\begin{figure*}[htp]
    \centering
    \includegraphics[width=0.9\textwidth]{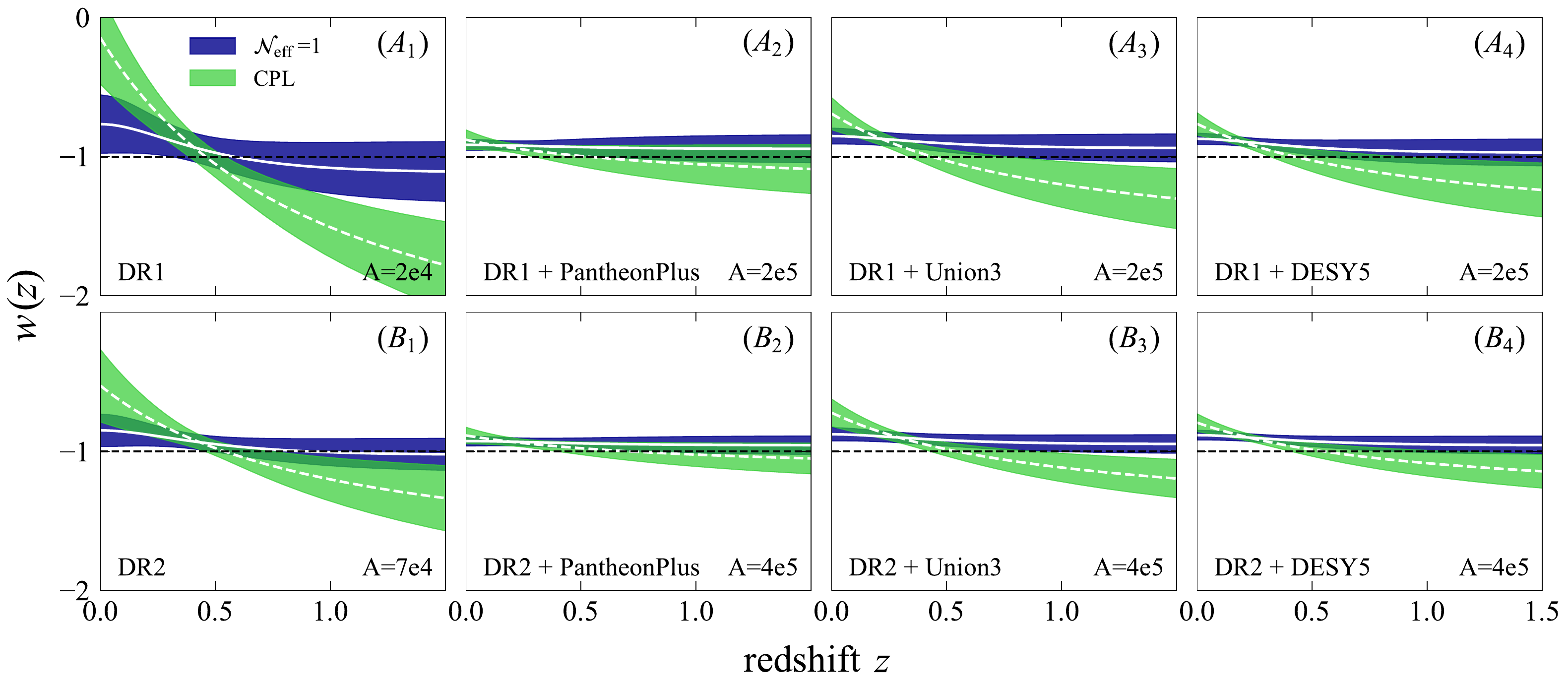}
    \includegraphics[width=0.9\textwidth]{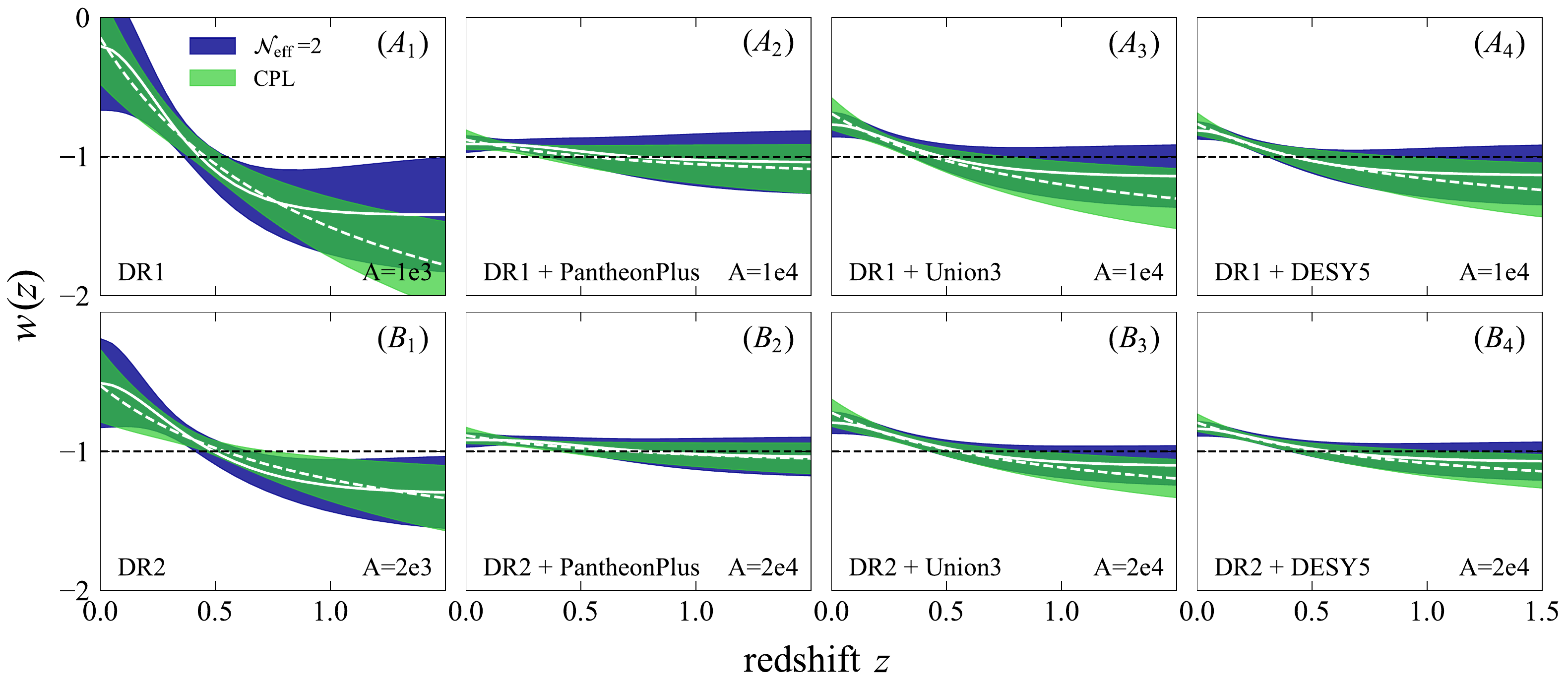}
    \caption{Same as Fig. \ref{fig:wz} in the main text, but the blue bands show the reconstruction result with $1$ (top) and $2$ (bottom) effective data modes by using enhanced correlation priors quantified by the $A$ factor. See texts for details.       
    }
    \label{fig:wz2}
\end{figure*}

\begin{figure*}
\includegraphics[width=0.9\textwidth]{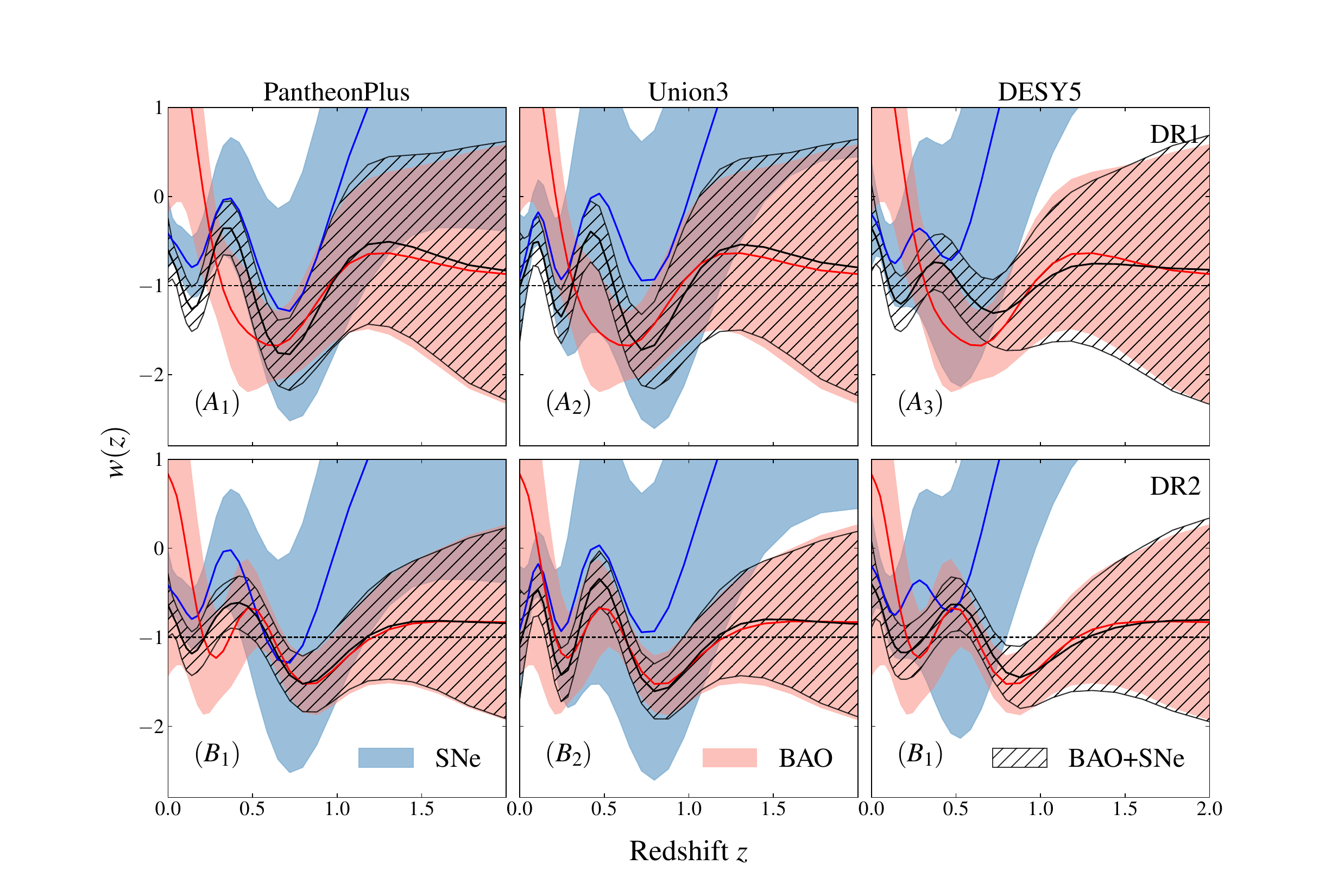}
    \caption{
        Reconstructed evolution of the dark energy equation-of-state parameter $w(z)$ 
        from SNe, BAO, and combined (SNe+BAO) data. 
        A BBN prior and a CMB measurement of $\theta_*$ are included when BAO measurements are used. 
        The top row ($A_1, A_2, A_3$) uses BAO data from DESI\,DR1, 
        while the bottom row ($B_1, B_2, B_3$) uses BAO data from DESI\,DR2. 
        From left to right, the columns show SNe data from PantheonPlus, Union3, and DESY5, respectively. 
        The blue and red shaded regions denote the 68\% confidence intervals from SNe-only 
        and BAO-only fits, respectively, 
        and the black hatched regions show the combined SNe+BAO constraints. 
        The black solid line in each panel indicates the best-fit reconstruction for the combined data. 
        The horizontal dotted line at $w=-1$ corresponds to the value for a cosmological constant.
    }
\label{fig:wz_sneonly}
\end{figure*}

\begin{figure*}
\includegraphics[width=0.9\textwidth]{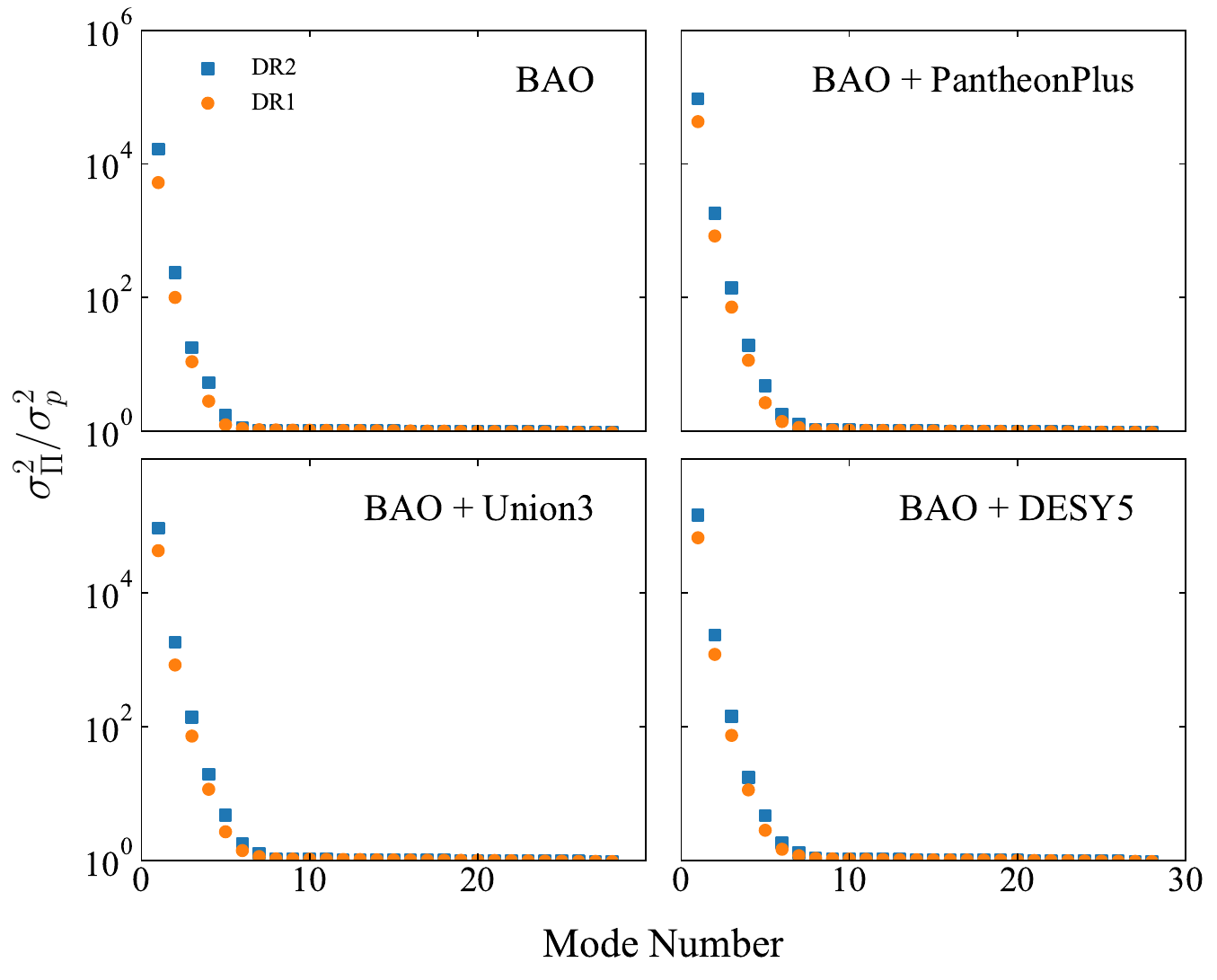}
    \caption{
        Ratio of the eigenvalues from the prior covariance matrix ($\mathbf{C}_\Pi$) to those from the posterior covariance matrix ($\mathbf{C}_p$) plotted against the covariant principal component analysis (CPCA) mode number. Each panel corresponds to a different data combination: Top left: BAO alone, Top right: BAO + PantheonPlus, Bottom left: BAO + Union3, and Bottom right: BAO + DESY5. Blue squares represent DESI DR2, while orange circles represent DESI DR1. The Big~Bang~Nucleosynthesis (BBN) priors and constraints on the CMB acoustic scale $\theta_*$ are imposed in all cases. Modes for which $\mathbf{C}_p$ has smaller eigenvalues than $\mathbf{C}_\Pi$ indicate tighter data constraints relative to the prior.
    }
\label{fig:CPCA}
\end{figure*}

In order to obtain the data modes without using any correlation prior, we need to set $A=0$. However, this makes the convergence of MCMC chains impractical. To overcome this problem, we use a Wiener filter projection.

Suppose a Fisher matrix of ${\bf w}$, denoted as ${\bf F}_{\rm data}$, can be obtained from a given dataset, and one performs two MCMC reconstructions to obtain ${\bf w}_{1}$ and ${\bf w}_{2}$ with two sets of modified prior covariance matrices $\tilde{\mathbf{C}}^{-1}_{\Pi,1}$ and  $\tilde{\mathbf{C}}^{-1}_{\Pi,2}$, respectively, then the following relation can be derived using the formulism of the Wiener filter projection \cite{Fables11}:
\begin{equation}\label{eq:wiener}
    {\bf w}_2 = \left({\bf I}+{\bf F}_{\rm data}^{-1} \tilde{{\bf C}}^{-1}_{\Pi,2}\right)^{-1} \left({\bf I}+{\bf F}_{\rm data}^{-1} \tilde{{\bf C}}^{-1}_{\Pi,1}\right) {\bf w}_1
\end{equation} where ${\bf I}$ is the identity matrix.

In practice, we set $A=0.01$ to perform a MCMC reconstruction to obtain ${\bf C}_{\rm p} (A=0.01)$, which is the posterior covariance matrix of ${\bf w}$, then we can obtain ${\bf F}_{\rm data}$ via\begin{equation}
 {\bf F}_{\rm data} = {\bf C}^{-1}_{\rm p} (A=0.01) - 0.01 {\bf C}^{-1}_{\Pi}.
\end{equation} The MCMC also returns the reconstructed ${\bf w}$, denoted as ${\bf w}_{A=0.01}$. Given ${\bf F}_{\rm data}$, ${\bf w}_1={\bf w}_{A=0.01}$ and $\smash{\tilde{{\bf C}}^{-1}_{\Pi,1}=0.01 {\bf C}^{-1}_{\Pi}}$, one can in principle reconstruct ${\bf w}$ for another modified correlation prior by a Wiener filter projection using Eq. (\ref{eq:wiener}).

One interesting special case is the  reconstruction ${\bf w}$ from data alone, without using any correlation prior. This is straightforward since ${\bf w}_{\rm data}$ can be found by setting $\tilde{\mathbf{C}}^{-1}_{\Pi,2}=0$, and the covariance matrix for ${\bf w}_{\rm data}$ is simply ${\bf F}^{-1}_{\rm data}$.

\section{Principal Components Analysis (PCA)} 

PCA is a widely used tool in cosmology to find uncorrelated eigenmodes by diagonalising the associated (inverse) covariance matrix. Specifically, \begin{equation}
    {\bf F}={\bf W}^T {\bf \Lambda} {\bf W}\, ,
\end{equation} where ${\bf F}$ is the inverse covariance matrix, ${\bf W}$ is the decorrelation matrix whose rows are the eigenvectors $e_i(z)$, and ${\bf \Lambda}$ is a diagonal matrix storing the eigenvalues $\lambda_i$ \cite{Huterer:2002hy}. In this work, we apply PCA for ${\bf F_{\rm data}}$ and $\tilde{\mathbf{C}}^{-1}_{\Pi}$, respectively, and show the eigenvalues and eigenvectors in Fig. \ref{fig:PCA}.

\section{Covariant Principal Components Analysis (CPCA)}

CPCA is a variant of PCA with nice features. It allows for finding eigenmodes for both the prior and posterior, therefore it is much easier to quantify the effective degrees of freedom in the system, especially when the prior is complex.

Unlike PCA, the CPCA decomposition is as in \cite{Raveri:2021dbu}:
\begin{equation}\label{eq:CPCA}
 {\bf C}_{\Pi} {\bf \Psi} ={\bf C}_{\rm p} {\bf \Psi} {\bf L}  \, , 
\end{equation} where the decomposition matrix $\bf \Psi$ and the diagonal matrix $\bf L$ store the CPCA eigenvectors and eigenvalues, respectively. The CPCA decomposition, Eq. (\ref{eq:CPCA}), yields\begin{eqnarray}
{\bf \Psi}^T {\bf C}_{\rm p} {\bf \Psi} &=&{\bf I},\nonumber\\
{\bf \Psi}^T {\bf C}_{\Pi} {\bf \Psi} &=&{\bf L}.
\end{eqnarray}

A ratio of eigenvalues of the prior and posterior matrices, presented in Supplementary Fig. \ref{fig:CPCA}, shows how informative the posterior modes are against the prior modes, and the effective number of degrees of freedom $\mathcal{N}_{\rm eff}$, can be calculated as
\begin{equation}
N_{\rm {eff}}=N-\operatorname{Tr}\left({\bf C}^{-1}_{\Pi}{\bf C}_{\rm p}\right),
\end{equation} where $N$ is the number of nominal parameters. The trace of the diagonal `improvement' matrix $\bf L$, denoted as $T\equiv {\rm Tr} (\bf L)$, quantifies the constraining power, and thus is a useful metric for a comparison among different datasets. Supplementary Table 4 shows $\mathcal{N}_{\rm eff}$, $T$ and the significance of deviation of $w(z)$CDM model from three models (in terms of the improved $\chi^2$) as illustrated in the table. Results are shown for four datasets.
\\

\section{Mock tests}
\label{sec:mock}

In this section, we show the reconstructed $w(z)$ from mock datasets, generated for models of $w=-1$, $w=-1.2$, $w(a)=-1.2+0.5(1-a)$ and $w(a)=-1.1-1.3(1-a)+11.2(1-a)^2-15.7(1-a)^3$, respectively, in Supplementary Figs \ref{fig:mock1} - \ref{fig:mock4}.

\begin{figure*}
\includegraphics[width=0.95\textwidth]{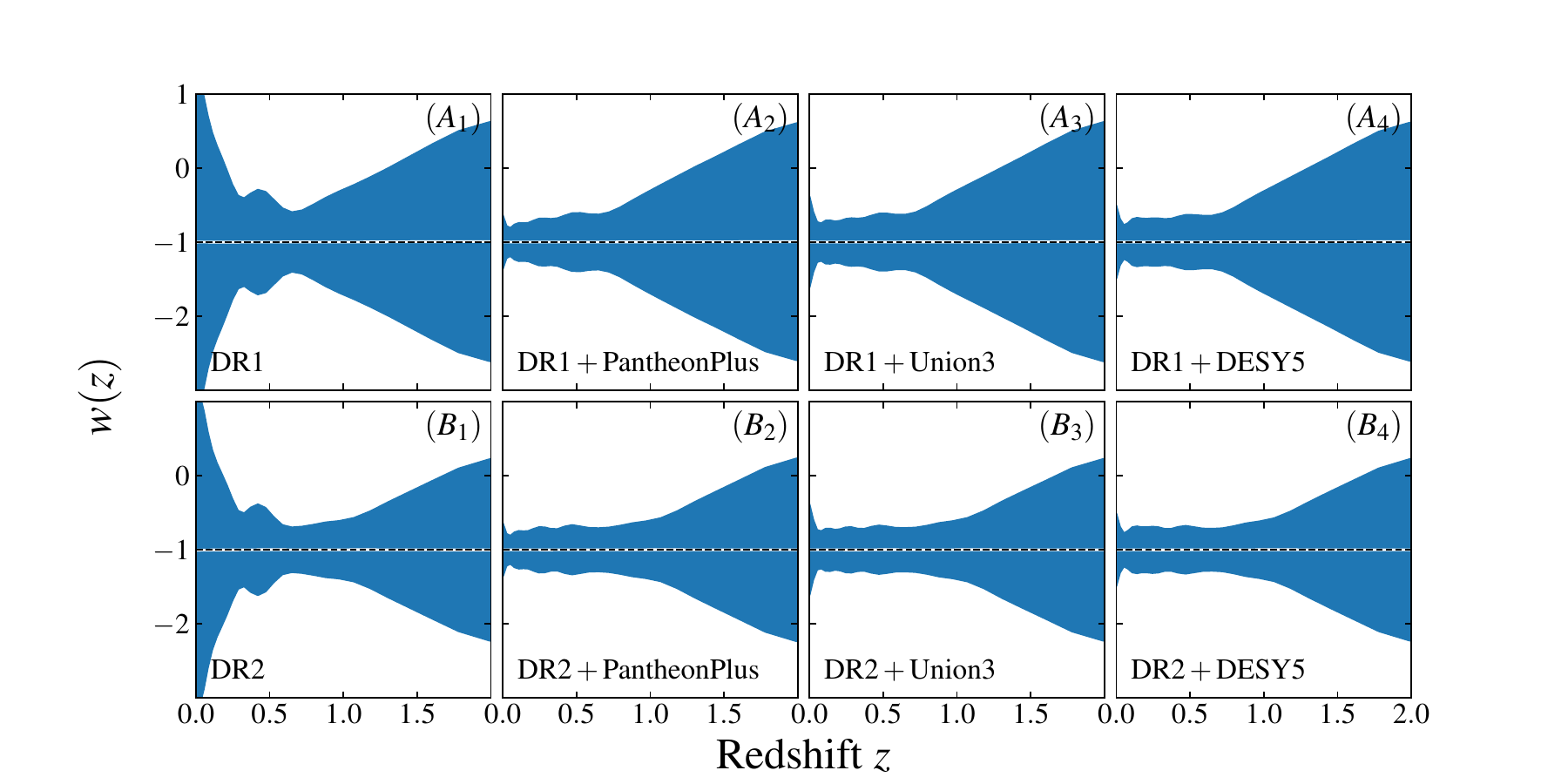}
    \caption{The best-fit (white solid curve) and 68\% uncertainty of $w(z)$ reconstructed from a mock dataset with the fiducial model of $w=-1$, as illustrated in black dashed line.        
    }
\label{fig:mock1}
\end{figure*}

\begin{figure*}
\includegraphics[width=0.95\textwidth]{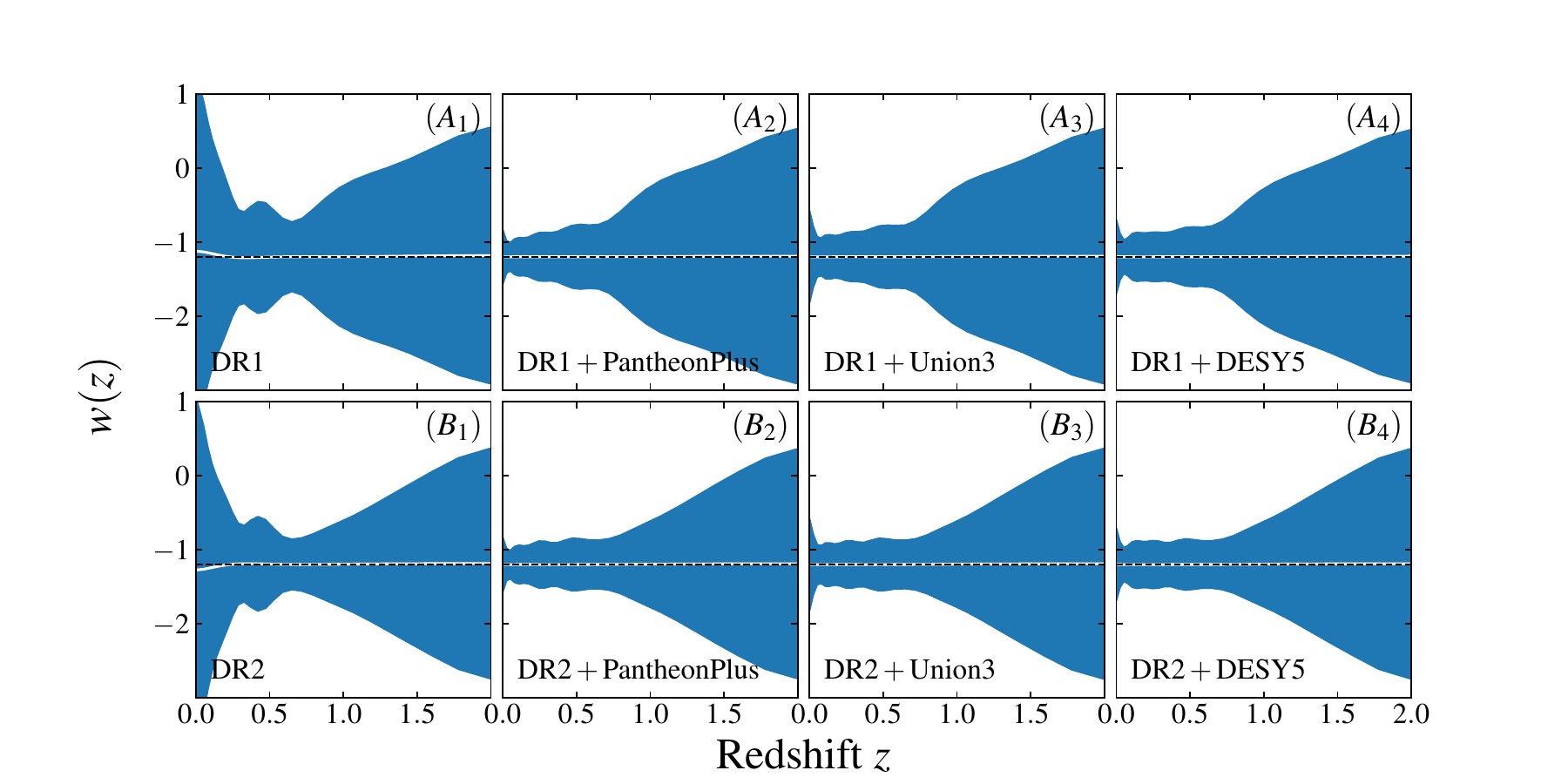}
    \caption{The best-fit (white solid curve) and 68\% uncertainty of $w(z)$ reconstructed from a mock dataset with the fiducial model of $w=-1.2$, as illustrated in black dashed line.        
    }
\label{fig:mock2}
\end{figure*}

\begin{figure*}
\includegraphics[width=0.95\textwidth]{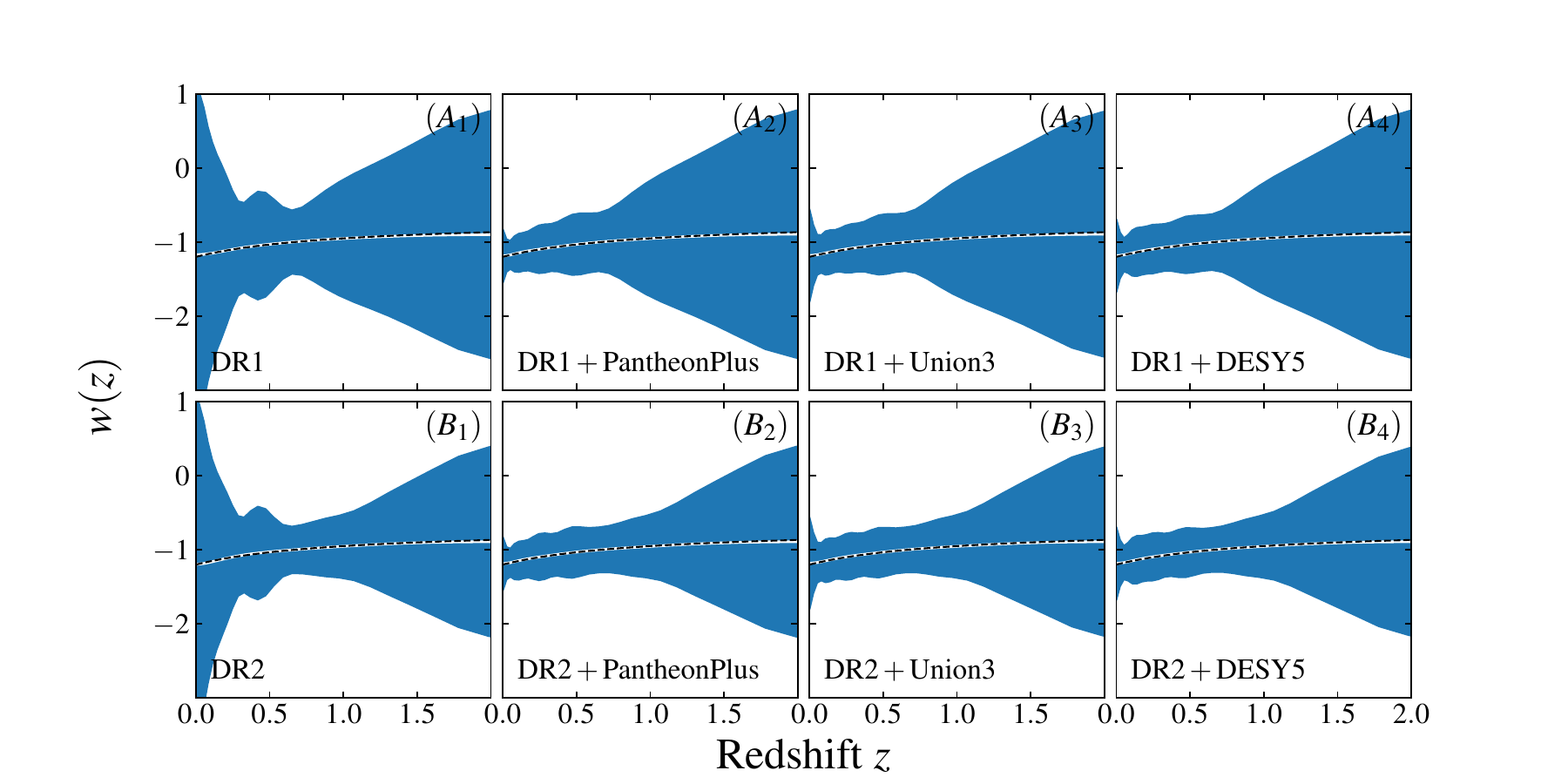}
    \caption{The best-fit (white solid curve) and 68\% uncertainty of $w(z)$ reconstructed from a mock dataset with the fiducial model of $w(a)=-1.2+0.5(1-a)$, as illustrated in black dashed line.        
    }
\label{fig:mock3}
\end{figure*}

\begin{figure*}
\includegraphics[width=0.95\textwidth]{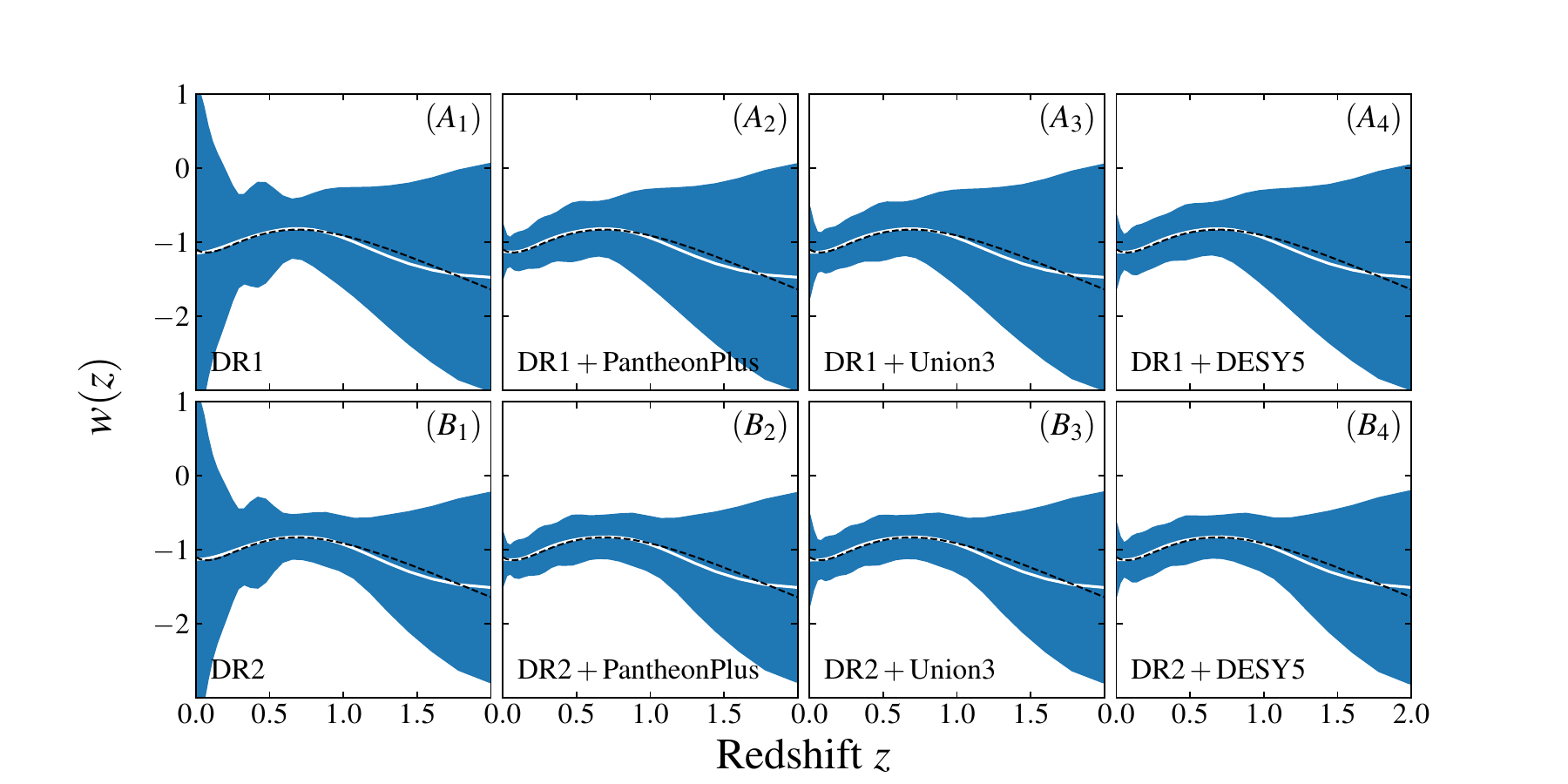}
    \caption{The best-fit (white solid curve) and 68\% uncertainty of $w(z)$ reconstructed from a mock dataset with the fiducial model of $w(a)=-1.1-1.3(1-a)+11.2(1-a)^2-15.7(1-a)^3$, as illustrated in black dashed line.        
    }
\label{fig:mock4}
\end{figure*}

\section{Supplementary Tables}

\begin{table*}[h!]
    \centering
    \renewcommand{\arraystretch}{1.2}
    \begin{tabular}{ll|ccc|ccc}
        \hline\hline       
        \multirow{2}{*}{Model} & \multirow{2}{*}{Dataset}  & \multicolumn{3}{c|}{BAO: DESI DR1} & \multicolumn{3}{c}{BAO: DESI DR2} \\
        \cline{3-8}
        & & $H_0$ [km s$^{-1}$Mpc$^{-1}$] & $\Omega_{\rm M}$ & $\Delta \chi^2$ & $H_0$ [km s$^{-1}$Mpc$^{-1}$] & $\Omega_{\rm M}$ & $\Delta \chi^2$ \\
        \hline
        \multirow{4}{*}{$\Lambda$CDM} & BAO & $68.69 \pm 0.63$ & $0.2934 \pm 0.0071$ & 0 & $68.53 \pm 0.48$ & $0.2955 \pm 0.0042$ & 0 \\
        & BAO + PantheonPlus & $68.30 \pm 0.61$ & $0.2989 \pm 0.0068$ & 0 & $68.41 \pm 0.48$ & $0.2974 \pm 0.0041$ & 0 \\
        & BAO + Union3 & $68.36 \pm 0.62$ & $0.2982 \pm 0.0070$ & 0 & $68.43 \pm 0.48$ & $0.2972 \pm 0.0042$ & 0 \\
        & BAO + DESY5 & $67.98 \pm 0.59$ & $0.3036 \pm 0.0067$ & 0 & $68.29 \pm 0.48$ & $0.2992 \pm 0.0042$ & 0 \\
        \hline
        \multirow{4}{*}{$w$CDM} & BAO & $68.6^{+1.8}_{-2.2}$ & $0.294^{+0.014}_{-0.013}$ & $-0.02$ & $67.7 \pm 1.1$ & $0.3004 \pm 0.0076$ & $-0.71$ \\
        & BAO + PantheonPlus & $67.22 \pm 0.82$ & $0.3030 \pm 0.0072$ & $-3.96$ & $67.22 \pm 0.72$ & $0.3037 \pm 0.0051$ & $-4.74$ \\
        & BAO + Union3 & $66.4 \pm 1.0$ & $0.3079 \pm 0.0083$ & $-5.23$ & $66.71 \pm 0.86$ & $0.3070 \pm 0.0061$ & $-5.60$ \\
        & BAO + DESY5 & $66.37 \pm 0.78$ & $0.3083 \pm 0.0071$ & $-10.79$ & $66.56 \pm 0.68$ & $0.3080 \pm 0.0050$ & $-11.66$ \\
        \hline
        \multirow{4}{*}{$w_0w_a$CDM} & BAO & $65.0^{+2.3}_{-3.7}$ & $0.337^{+0.041}_{-0.029}$ & $-1.79$ & $64.2^{+1.9}_{-2.5}$ & $0.343^{+0.028}_{-0.025}$ & $-3.37$ \\
        & BAO + PantheonPlus & $67.55 \pm 0.87$ & $0.3053 \pm 0.0076$ & $-4.80$ & $67.26 \pm 0.72$ & $0.3073 \pm 0.0061$ & $-5.86$ \\
        & BAO + Union3 & $66.2 \pm 1.0$ & $0.322 \pm 0.011$ & $-9.23$ & $65.90 \pm 0.92$ & $0.3233 \pm 0.0096$ & $-10.62$ \\
        & BAO + DESY5 & $66.92 \pm 0.77$ & $0.3148 \pm 0.0076$ & $-14.42$ & $66.59 \pm 0.67$ & $0.3157 \pm 0.0061$ & $-16.16$ \\
        \hline
        \multirow{4}{*}{$w(z)$CDM} & BAO & $61.3^{+5.6}_{-17}$ & $0.43^{+0.15}_{-0.21}$ & $-6.81$ & $65.2^{+6.9}_{-18}$ & $0.373^{+0.091}_{-0.21}$ & $-6.98$ \\
        & BAO + PantheonPlus & $67.2 \pm 1.0$ & $0.3132 \pm 0.0097$ & $-15.98$ & $66.90 \pm 0.86$ & $0.3141 \pm 0.0076$ & $-13.80$ \\
        & BAO + Union3 & $66.6 \pm 1.7$ & $0.319 \pm 0.016$ & $-18.03$ & $66.4 \pm 1.6$ & $0.319 \pm 0.015$ & $-20.07$ \\
        & BAO + DESY5 & $66.3 \pm 1.1$ & $0.3213 \pm 0.011$ & $-17.70$ & $66.05 \pm 0.96$ & $0.3226 \pm 0.0089$ & $-21.33$ \\
        \hline\hline
    \end{tabular}
\caption{Mean values and 68\% confidence-level (CL) uncertainties for the Hubble constant ($H_0$) and the present-day matter density parameter ($\Omega_{\rm M}$) extracted under four different cosmological models (listed in the first column) and four data combinations (BAO alone and BAO combined with each of the three supernova datasets). For non-Gaussian posterior distributions, we provide asymmetric error bars. The final column gives the improvement in $\chi^2$ relative to the corresponding $\Lambda$CDM fit. Results are shown for DESI Data Release~1 (DR1) and Data Release~2 (DR2) BAO measurements, enabling a direct comparison of constraints at different survey stages.
}
    \label{tab:params}
\end{table*}

\begin{table*}[]
\begin{center}
{
\begin{tabular}{c|cc|cc}
\hline\hline
& \multicolumn{2}{c|}{BAO: DESI DR1} & \multicolumn{2}{c}{BAO: DESI DR2}\\
\hline
Model/Dataset &  $w$ or $w_{0}$ &  $w_{a}$  &  $w$ or $w_{0}$ &  $w_{a}$   \\
\hline
{\bf $w$CDM} &  &   \\
BAO& $-0.997^{+0.093}_{-0.079}$ & -  & $-0.964\pm0.047$ & -  \\
BAO + PantheonPlus& $-0.937\pm0.032$ & -   & $-0.940\pm0.027$ & -   \\
BAO + Union3& $-0.902\pm0.044$ & - & $-0.918\pm0.034$ & - \\
BAO + DESY5& $-0.899\pm0.031$ & - & $-0.912\pm0.026$ & - \\
\hline
{\bf $w_0 w_a$CDM} &  &   \\
BAO& $-0.54^{+0.43}_{-0.23}$ & $<-1.03$  & $-0.52^{+0.29}_{-0.25}$ & $-1.40^{+0.84}_{-0.88}$  \\
BAO + PantheonPlus& $-0.874\pm0.071$ & $-0.40\pm0.39$   & $-0.883\pm0.059$ & $-0.30\pm0.27$   \\
BAO + Union3& $-0.68\pm0.12$ & $-1.10\pm0.55$ & $-0.713\pm0.098$ & $-0.83\pm0.37$ \\
BAO + DESY5& $-0.755\pm0.080$ & $-0.86\pm0.44$ & $-0.790\pm0.063$ & $-0.61\pm0.29$ \\
\hline\hline
\end{tabular}}
\caption{Mean values of the dark energy equation-of-state parameters for \(\Lambda\text{CDM}\) extensions (i.e.\ \(w\text{CDM}\) and \(w_0 w_a \text{CDM}\)) derived from the indicated BAO datasets (DESI DR1 and DR2). Each row presents results for a different data combination. The quoted uncertainties correspond to 68\% confidence intervals.}
\label{table:w0wa}
\end{center}
\end{table*}

\begin{table*}[h!]
    \centering
    \renewcommand{\arraystretch}{1.2}
    \begin{tabular}{c|cccc|cccc}
        \hline\hline
        $\mathcal{N}_{\rm eff}$ & DR1 & DR1+PantheonPlus & DR1+Union3 & DR1+DESY5 & DR2 & DR2+PantheonPlus & DR2+Union3 & DR2+DESY5 \\
        \hline
         1 & $0.0 \pm 0.6$ & $1.0 \pm 0.6$ & $1.7 \pm 0.6$ & $3.2 \pm 0.6$ & $0.3 \pm 0.6$ & $1.0 \pm 0.6$ & $2.3 \pm 0.6$ & $3.8 \pm 0.6$ \\
         2 & $0.2 \pm 0.7$ & $1.5 \pm 0.6$ & $2.5 \pm 0.6$ & $4.1 \pm 0.6$ & $0.4 \pm 0.7$ & $1.3 \pm 0.6$ & $3.0 \pm 0.6$ & $5.0 \pm 0.6$ \\
         3 & $0.0 \pm 0.7$ & $1.5 \pm 0.7$ & $2.8 \pm 0.7$ & $3.4 \pm 0.7$ & $-0.5 \pm 0.7$ & $1.4 \pm 0.7$ & $3.3 \pm 0.7$ & $5.2 \pm 0.7$ \\
         4 & $-1.2 \pm 0.8$ & $1.1 \pm 0.7$ & $2.3 \pm 0.7$ & $2.4 \pm 0.7$ & $-1.4 \pm 0.7$ & $1.1 \pm 0.7$ & $3.1 \pm 0.7$ & $4.7 \pm 0.7$ \\
         5 & $-$ & $-0.1 \pm 0.7$ & $0.7 \pm 0.7$ & $1.2 \pm 0.7$ & $-$ & $-0.3 \pm 0.7$ & $2.1 \pm 0.7$ & $3.0 \pm 0.7$ \\
         6 & $-$ & $-5.2 \pm 0.7$ & $-4.2 \pm 0.7$ & $-5.1 \pm 0.7$ & $-$ & $-3.0 \pm 0.7$ & $-0.6 \pm 0.7$ & $-0.8 \pm 0.7$ \\
        \hline
         1 & $1.1$ & $1.7$ & $2.2$ & $2.9$ & $1.2$ & $1.8$ & $2.0$ & $2.9$ \\
         2 & $1.7$ & $2.3$ & $2.8$ & $3.5$ & $1.8$ & $2.4$ & $2.8$ & $3.6$ \\
         3 & $2.2$ & $2.8$ & $3.2$ & $3.8$ & $2.2$ & $2.8$ & $3.2$ & $3.9$ \\
         4 & $2.4$ & $3.2$ & $3.5$ & $4.0$ & $2.4$ & $3.1$ & $3.6$ & $4.2$ \\
         5 & $-$ & $3.6$ & $3.8$ & $4.1$ & $-$ & $3.3$ & $3.9$ & $4.3$ \\
         6 & $-$ & $3.9$ & $3.9$ & $4.2$ & $-$ & $3.6$ & $4.1$ & $4.5$ \\
        \hline\hline
    \end{tabular}
    \caption{Comparison of $\Delta \ln E$ (numbers with error bars in the upper half of the table) and the signal-to-noise ratio (SNR) for detecting deviations from $w=-1$ (numbers in the lower half of the table) across different values of $\mathcal{N}_{\rm eff}$ using four data combinations. The left half of the table corresponds to DESI DR1 results, while the right half presents the DESI DR2 results.}
    \label{tab:evidence}
\end{table*}

\begin{table*}[h!]
\begin{center}
\begin{tabular}{c|c|c|c|c|c}
\hline\hline
\multirow{2}{*}{Dataset} 
 & \multirow{2}{*}{$\mathcal{N}_{\rm eff}$} & \multirow{2}{*}{$T$} & SNR of & SNR of & SNR of \\
  & & &$w\ne-1$&$w\ne~$constant&$w\ne w_0+w_a(1-a)$\\
\hline
DR1        &$4.9$ &$5394.7$   &$2.8$  &$2.8$ & $2.1$\\
DR1 + PantheonPlus  &$6.2$ &$61373.7$  &$3.8$  &$3.1$ & $3.1$\\ 
DR1 + Union3   &$5.9$ &$43855.4$  & $3.9$ &$3.0$ & $2.3$\\
DR1 + DESY5  &$6.1$ & $68266.3$ &$4.2$  &$2.4$ & $1.6$\\
\hline
DR2        &$5.2$ &$16963.5$   &$2.6$  &$2.3$ & $1.6$\\
DR2 + PantheonPlus  &$6.7$ &$155945.6$  &$3.6$  &$2.8$ & $2.7$\\ 
DR2 + Union3   &$6.4$ &$95745.0$  & $4.2$ &$3.3$ & $2.6$\\
DR2 + DESY5  &$6.5$ & $148540.7$ &$4.5$  &$2.8$ & $2.1$\\
\hline\hline
\end{tabular}
\caption{Effective degrees of freedom $\mathcal{N}_{\rm eff}$, trace $T$, and the signal-to-noise ratio (SNR) of deviations of the $w(z)$CDM model from three reference models---$\Lambda$CDM, $w$CDM, and $w_0w_a$CDM---across four data combinations. The top section lists DESI Data Release~1 (DR1) results, while the bottom section shows DESI Data Release~2 (DR2).}
\label{table:CPCA}
\end{center}
\end{table*}

\begin{table*}[h!]
\begin{center}
{
\begin{tabular}{c|c|c|c|c|c}
\hline\hline
Dataset & $\gamma_1$ & $\gamma_2$ & $\gamma_3$ & $\gamma_4$ & $\gamma_5$\\
\hline
DR1        &$0.23 \pm 0.40$ &$-1.86 \pm 0.78$   &$0.3 \pm 1.3$ &$1.6 \pm 1.7$   &$0.7 \pm2.6$ \\
DR1 + PantheonPlus  &$0.28 \pm 0.12$ &$0.16 \pm 0.25$   &$-0.51 \pm 0.43$ &$-1.66 \pm 0.65$   &$0.92 \pm0.87$ \\ 
DR1 + Union3   &$0.52 \pm 0.18$ &$-0.37 \pm 0.32$   &$-0.14 \pm 0.52$ &$-0.95 \pm 0.73$   &$2.7 \pm1.1$ \\
DR1 + DESY5  &$0.42 \pm 0.12$ &$-0.30 \pm 0.25$   &$-0.17 \pm 0.49$ &$-1.22 \pm 0.73$ &$0.1 \pm 1.1$ \\
\hline\
DR2        &$0.28 \pm 0.21$ &$-0.90 \pm 0.47$   &$-0.85 \pm 0.73$ &$0.72 \pm 0.90$   &$-0.3 \pm2.0$ \\
DR2 + PantheonPlus  &$0.26 \pm 0.11$ &$0.15 \pm 0.20$   &$0.01 \pm 0.33$ &$-1.21 \pm 0.48$   &$0.56 \pm 0.70$ \\ 
DR2 + Union3  &$0.44 \pm 0.15$ &$-0.36 \pm 0.25$   &$-0.08 \pm0.39$ &$-1.23 \pm 0.58$   &$0.93 \pm0.73$ \\
DR2 + DESY5  &$0.41 \pm 0.11$ &$-0.16 \pm 0.20$   &$-0.22 \pm 0.36$ &$-1.35 \pm 0.59$ &$0.38 \pm 0.74$ \\
\hline\hline
\end{tabular}}
\caption{The amplitude of the first five PCA eigenmodes of $w(z)$ reconstructed from various data combinations.
}
\label{table:PCA}
\end{center}
\end{table*}

\begin{table*}[h!]
\begin{center}
{
\begin{tabular}{c|c|c|c|c}
\hline\hline
\multirow{2}{*}{Dataset} 
  & $\Delta \chi^2$& $\Delta \chi^2$ & $\Delta \chi^2$ & $\Delta \chi^2$ \\
  &($\mathcal{N}_{\rm eff}$=1)&($\mathcal{N}_{\rm eff}$=2) &($w_0 w_a$CDM) &($w(z)$CDM)\\
\hline
DR1        &$-1.34$ &$-4.04$ &$-6.98$ &$-8.01$  \\
DR1 + PantheonPlus  &$-5.93$ &$-6.20$ &$-4.78$ &$-14.81$ \\ 
DR1 + Union3   &$-7.51$ &$-9.45$ &$-9.00$ &$-14.93$ \\
DR1 + DESY5  &$-12.06$ & $-14.14$ &$-14.25$ &$-17.51$\\
\hline
DR2        &$-2.33$ &$-5.12$ &$-3.57$ &$-6.61$  \\
DR2 + PantheonPlus  &$-5.79$ &$-6.30$ &$-5.79$ &$-13.29$ \\ 
DR2 + Union3   &$-7.78$ &$-10.26$ &$-9.97$ &$-17.64$ \\
DR2 + DESY5  &$-13.71$ & $-15.68$ &$-16.24$ &$-20.34$\\
\hline\hline
\end{tabular}}
\caption{The improved $\chi^2$ of four different dark energy models with respect to the $\Lambda$CDM model derived from various data combinations.}
\label{table:delta_chi2}
\end{center}
\end{table*}

\end{document}